\immediate\write18{makeindex \jobname.nlo -s nomencl.ist -o \jobname.nls}

\documentclass[final,3p,times,authoryear]{elsarticle}

\usepackage{graphicx}
\usepackage{verbatim}
\usepackage[figuresright]{rotating}
\usepackage{savesym}
\savesymbol{iint}
\usepackage{txfonts}
\restoresymbol{TXF}{iint}
\usepackage{amssymb}
\usepackage{amsfonts}
\usepackage{mathrsfs}
\usepackage{adjustbox}
\usepackage{mdframed}
\usepackage{eurosym}
\usepackage{lipsum}
\usepackage{smartdiagram}
\usepackage{dutchcal}
\usepackage{algorithm}
\usepackage[noend]{algpseudocode}

\graphicspath{ {} }

\usepackage{chngcntr}
\usepackage{tabularx}
\usepackage{booktabs}
\usepackage{lscape}
\usepackage{pifont}
\usepackage{siunitx}
\usepackage{eurosym}
\usepackage[allcolors=blue]{hyperref}
\usetikzlibrary{arrows.meta}

\usetikzlibrary{shadows.blur}
\usetikzlibrary{positioning}
\usepackage{standalone}
\usepackage{makecell}
\usepackage{arydshln}
\usepackage{array}

\usepackage{nomencl}
\usepackage{etoolbox}
\usepackage{wrapfig}
\usepackage{framed} 
\usepackage[skins,breakable, most]{tcolorbox}

\renewcommand*\nompreamble{\begin{multicols}{2}}
\makeatletter
\def\ps@pprintTitle{%
 \let\@oddhead\@empty
 \let\@evenhead\@empty
 \def\@oddfoot{}%
 \let\@evenfoot\@oddfoot}
\makeatother

\usepackage{epsfig}
\usepackage{epstopdf} 
\usepackage{todonotes}
\usepackage{setspace}
\usepackage{acronym}
\usepackage{siunitx}
\usepackage[utf8x]{inputenc} 
\usepackage{stfloats}

\usepackage[caption = false]{subfig}

\usepackage{booktabs}

\makeatletter
\newcommand{\myonecolumn}{\clearpage \global \columnwidth \textwidth \global \hsize \columnwidth \global \linewidth \columnwidth \global \@twocolumnfalse \col@number \@ne \@floatplacement}
\makeatother

\usepackage{lmodern,textcomp}
\usepackage{rotating} 
\usepackage{multicol} 
\usepackage{multirow} 
\usepackage{mathtools, cuted}
\usepackage{lipsum}
\usepackage{tabularx} 
\newcolumntype{L}[1]{>{\raggedright\arraybackslash}p{#1}} 
\newcolumntype{C}[1]{>{\centering\arraybackslash}p{#1}} 
\newcolumntype{R}[1]{>{\raggedleft\arraybackslash}p{#1}} 
\usepackage{amssymb}
\usepackage{pifont}
\RequirePackage[ngerman=ngerman-x-latest]{hyphsubst}
\usepackage{microtype}
\usepackage[american]{babel} 
\usepackage{float}

\usepackage{lineno}

\biboptions{square,comma,sort&compress}

\setcitestyle{numbers}
\journal{Applied Energy}

\usepackage{tikz}
\usetikzlibrary{arrows,shapes,snakes,automata,backgrounds,petri, positioning,chains,decorations.pathmorphing}
\usetikzlibrary{matrix,calc,chains,arrows,shapes}
\begin{document}

\title{Competition between simultaneous demand-side flexibility options: The case of community electricity storage systems}

\author[DTU,IIRM]{Fabian Scheller\corref{cor1}}
\ead{fjosc@dtu.dk}
\author[IIRM]{Robert Burkhardt}
\author[IIRM]{Robert Schwarzeit}
\author[DTU]{Russell McKenna}
\author[IIRM]{Thomas Bruckner}
\cortext[cor1]{Corresponding author}

\address[DTU]{Energy Systems Analysis, Division of Sustainability, Department of Technology, Management and Economics, Technical University of Denmark (DTU)}
\address[IIRM]{Institute for Infrastructure and Resources Management (IIRM), University Leipzig}

\begin{abstract}
Community electricity storage systems for multiple applications promise benefits over household electricity storage systems. More economical flexibility options such as demand response and sector coupling might reduce the market size for storage facilities. This paper assesses the economic performance of community electricity storage systems by taking competitive flexibility options into account. For this purpose, an actor-related, scenario-based optimization framework is applied. The results are in line with the literature and show that community storage systems are economically more efficient than household storage systems. Relative storage capacity reductions of community storage systems over household storage systems are possible, as the demand and generation profiles are balanced out among end users. On average, storage capacity reductions of 9\% per household are possible in the base case, resulting in lower specific investments. The simultaneous application of demand-side flexibility options such as sector coupling and demand response enable a further capacity reduction of the community storage size by up to 23\%. At the same time, the competition between flexibility options leads to smaller benefits regarding the community storage flexibility potential, which reduces the market viability for these applications. In the worst case, the cannibalization effects reach up to 38\% between the flexibility measures. The losses of the flexibility benefits outweigh the savings of the capacity reduction whereby sector coupling constitutes a far greater influencing factor than demand response. Overall, in consideration of the stated cost trends, the economies of scale, and the reduction possibilities, a profitable community storage model might be reached between 2025 and 2035. Future work should focus on the analysis of policy frameworks.
\end{abstract}


\maketitle

\section*{Highlights}
\begin{itemize}
\item Community storage systems are conceptualized as complex techno-economic system.
\item Optimization of the competition between simultaneous residential flexibility options.
\item Losses due to simultaneous application of flexibility options outweigh the benefits.
\item Sector coupling constitutes a greater competition factor than demand response.
\item Community storage systems will become economically profitable in the next few years.
\end{itemize}

\section{Introductory remarks}
\label{Introduction}

\subsection{Research context}
As part of an increased diffusion of decentralized renewable energy (RE) technologies, an additional need for flexibility arises. One key technology to compensate the increasing intermittency in electricity generation are battery energy storage (BES) systems. BES systems have the capability to improve the dispatchability of RE technologies \cite{PARRA2017} and provide additional flexibility which ensures the resilience and reliability of the existing power system \cite{TERLOUW2019356}. Since various RE sources are installed at the household level, there is a growing interest for BES systems that are located in spatial proximity. 

Nowadays, electric batteries are often employed as household electricity storage (HES) systems that are installed in single dwellings and used by prosumers to maximize their share of individual self-consumption. However, the economic concerns due to the high costs of Lithium-ion (Li-ion) batteries remain a major barrier preventing their widespread implementation \cite{mckinsweb,yang.2018}. Recent studies indicate that BES systems for multiple application areas promise higher economic and ecological advantages \citep{IWES2014,Zeh2015,braeuer2019battery,dong2020techno}. This seems to be decisive since more economical community flexibility options (e.g. sector coupling in terms of power-to-heat or electric vehicles, and smart appliances with the ability to shift load) might further reduce the market size for BES applications \citep{phdparra}. Community energy storage (CES) systems, which are battery storage systems shared by prosumers at the community level, represent a potential business model for an energy storage with multi-use function. 

Some techno-economic advantages of CES systems acknowledged by previous research include lower specific investments due to economies of scale \cite{VANDERSTELT2018266,PARRA2017aop}, and relative storage capacity reduction of CES over HES systems as the demand and generation profiles of CES-connected households are balanced out before storing is necessary 
\cite{PARRA2017aop,PARRA2015}. Furthermore, a CES can theoretically serve multiple applications simultaneously rather than offering one service only and thus capture additional revenue streams  \cite{LOMBARDI2017,phdmetz}. The aforementioned techno-economic advantages potentially contribute to lowering the profitability threshold for CES systems. 

\subsection{Related work}
Understanding the value of CES systems is complex and remains a major challenge \cite{KOIRALA2019209}. Since further flexibility options compete with CES solutions, this review focuses on literature which assess the economic viability of CES systems by taking into account the capacity reduction potential but also competing flexibility options. A summary is given in Table \ref{zusammenfassung}. 

First, the reviewed literature is classified by the method used to assess the economic performance and optimal size. All reviewed studies use mathematical models which can be further classified into simulation models (Sim.), optimization models (Opt.) or a hybrid of both. Second, different metrics for assessing the techno-economic performance are proposed. In terms of economics, an established set of indicators are widely used including the levelized cost of energy storage (LCOES), the levelized value of energy storage (LVOES), payback period (PBP), net present value (NPV) and the internal rate of return (IRR), as well as the total benefit of a storage in terms of the equivalent annual value (EAV). Additionally, common performance indicators have been the absolute self-consumption (SC) or the SC ratio (SCR). Other valid metrics constitute the round-trip efficiency ($\eta$) or the equivalent full cycles (EFC). Third, particular attention is paid to whether the study integrates flexibility options in the assessment. Demand response (DR), heat pumps (HPs) as well as battery electric vehicles (BEVs) have been included. Fourth, the studies differ regarding the combination of application areas. The CES applications are divided into end user and operator services and are concerned with maximizing self-consumption (MSC), bill management (BM), spot market arbitrage (ARB), and peak shaving (PS). Some studies also show the potential of CES in participating in balancing markets such as frequency control application (FR) to gain additional revenues. Fifth, the overview contains the assumed investments. Sixth, the table states if the CES business case constitutes an economically feasible business case. All further results are shortly summarized in the far-right column. 

\newpage
\setlength\dashlinedash{0.2pt}
\setlength\dashlinegap{1.5pt}
\setlength\arrayrulewidth{0.3pt}
\begin{landscape}
\begin{table}[h]
\caption[Overview of studies reviewed]{Overview of reviewed studies investigating the techno-economic performance of HES and CES systems}
\centering
\scalebox{0.8}{
\begin{tabular}{@{} l c c c c c  c c c c c @{}}
    \toprule
    \textbf{Ref.} & Method & \multicolumn{2}{c}{Key assessment indicators}&  \multicolumn{2}{c}{Appl. flexibility options} & \multicolumn{2}{c}{CES application}  & Assumed & Econ. & Main\\
    \cmidrule(lr){3-4}  \cmidrule(lr){5-6} \cmidrule(lr){7-8}
    & &Performance & Economic&DR&Sector coupling & End user & Operator  &CAPEX &viable & results \\
    \midrule
      \textbf{\cite{TERLOUW2019356}}& Opt.& SCR & LCOES, PBP & - & (HP), BEV& -& ARB, PS & 759 - 2,352\EUR{}/kWh  &\ding{51} & \makecell{Combination of ARB+PS for CES \\is promising business case;   Li-ion\\ is best performing  technology}\\ \hdashline \addlinespace[0.2cm] 
   \textbf{\cite{BARBOUR2018} }& Sim./Opt.&- & EAV, IRR, NPV & - & -& MSC & -& \makecell{750\$/kWh\\+10\$/kW p.a.}  &\ding{51} &\makecell{Optimum CES capacity \\is 65\% of HES; IRR of \\9.3\% (CES) vs. 8.0\% (HES)}\\ \hdashline \addlinespace[0.2cm]
                
    \textbf{\cite{VANDERSTELT2018266}} &Opt.& SC, SCR &LCOES, PBP&\ding{51}&(HP)&MSC, BM&-& 1,000 \EUR/kWh& \ding{55}&\makecell{MSC\textgreater BM; op. cost improvement \\ of 22\% (HES) and 30\%(CES);\\SC improvement of \\23\%(HES) and 29\%(CES) }\\ \hdashline \addlinespace[0.2cm]
    \textbf{\cite{KLINGLER20181064}} &Opt.& SC&EAV&-&HP, BEV&MSC&-& 907 \EUR/kWh (2015)& \ding{51}&\makecell{SC\textsubscript{HES} decreased by 35\% with HP \\ and increased with BEV by 9\%;\\corresponding savings decreased \\by 35\% and increased by 17\% }\\ \hdashline \addlinespace[0.2cm]
     \textbf{\cite{PARRA2017aop}} & Sim./Opt. & $\eta$, EFC& LCOES, LVOES, IRR& -&(HP)& MSC, BM&-&1,185 \pounds/kWh (2020)& \ding{55} &\makecell{CES\textgreater  HES; MSC and  BM are\\ competing applications for CES}   \\ \hdashline \addlinespace[0.2cm]
    \textbf{\cite{PARRA2015}} & Sim./Opt. & $\eta$, EFC& LCOES, LVOES, IRR& -&(HP)&MSC &-&1,185 \pounds/kWh (2020)& \ding{55}  &\makecell{Cost reduction of 37\% in 2020 \\ due to community approach \\(instead of single home HES)}  \\ \hdashline \addlinespace[0.2cm]
    \textbf{\cite{PARRA2016}} & Sim./Opt. & $\eta$, EFC& LCOES, LVOES, IRR& -&HP& BM&-&1,185 \pounds/kWh (2020)& \ding{55}  &\makecell{HP penetration of 14\% reduces  \\ LCOES by 5-8\% and increases \\ CES profitability; LCOES reduced\\  by 56\% changing from HES to CES}  \\ \hdashline \addlinespace[0.2cm]
    \textbf{\cite{Panagiotou}} & Sim. & -& Operational cost& -&-&MSC, BM&-& -&- &\makecell{Operational energy costs are \\reduced by 31\% with CES; CES\\  enables additional revenues }  \\ \hdashline \addlinespace[0.2cm]
    \textbf{\cite{ARGHANDEH2014}} & Sim./Opt. & -& Operational cost& -&-&-&ARB& -&-  &\makecell{Operational energy costs are \\ reduced up to 65\% when \\ using a CES system} \\ 
    \addlinespace[0.2cm]
    \textbf{\cite{telaretti2016economic}} & Sim. & -& Cash flow, NPV, IRR & -&-&BM&-& 319 - 2,074 \EUR/kWh (2015)&\ding{55} &\makecell{Li-ion is best performing   \\ technology and will become \\  cost effective in 2020} \\ 
    \bottomrule
\end{tabular}}
\label{zusammenfassung}
\end{table}
\end{landscape}
\clearpage

Investments are seen as one of the major drawbacks to deliver viable business cases \cite{VANDERSTELT2018266,PARRA2017aop,sardi2017}. Due to economies of scale, larger CES systems can be realized with lower specific investments than smaller HES systems  \cite{ISEA2018,PARRA2015,Sieg2015,IWES2014}. Considering this, \cite{PARRA2015} assume manufacturing economies of scale for systems larger a certain threshold.  

Furthermore, \cite{VANDERSTELT2018266}, \cite{Schaffler2015} and \cite{BARBOUR2018} identified capacity reduction potentials between 24\% and 35\% for CES systems compared to HES systems without abandoning any monetary benefits. This means that the same savings and revenues of HES systems can be realized with CES systems while having a significantly lower storage capacity. In total, the studies demonstrate that CES systems generally exhibit a better economic performance than HES systems. This is justified by economies of scale, the integration of multiple applications as well as the aggregation of demand in the community approach. 

At the same time, the reviewed studies differ widely regarding decisive evaluation parameters such as integrated flexibility options. The different assumptions lead to different, even contradicting findings at first glance. While \cite{KLINGLER20181064} states that HPs increase the profitability of CES systems, \cite{PARRA2016} have opposing findings with numerical evidence. A closer inspection, however, reveals that different application purposes for CES systems may constitute an explanation for the contradictory findings. In \cite{KLINGLER20181064} the battery is solely used for increasing the on-site consumption of photovoltaic (PV)-generated electricity and value by the HES is created by storing excess PV-generated electricity for later use. When a HP is introduced, electricity for storing can be directly used by the HP resulting in a lower utilization of the HES and consequently a lower economic benefit of the battery. Whereas in \cite{PARRA2016}, the CES is used for BM and financial benefits are derived when charging the battery in times of favorable electricity tariffs. The presence of a HP increases the total electricity load and a higher share can be met with off-peak electricity. Beyond that, CES systems currently do not represent the cheapest option for community flexibility \citep{Agora2014,IWES2014}. Thus, prior to large-scale use of storage systems, load management and the development of cross-sectoral flexibility options are more applicable \citep{phdparra}. 

In summary, the techno-economic performance of CES systems has been investigated on the basis of different assumptions and various applications as summarized in Table \ref{zusammenfassung}. Nevertheless, there is a lack of knowledge about the effect of emerging flexibility options on profitability and sizing of CES systems and thus the market viability of these systems. Given that the simultaneous application of more economical flexibility options like demand response and sector coupling might reduce the optimal CES capacity, which leads to lower storage-related investments, they also might cannibalize the flexibility potential of the CES system, which leads to smaller storage-related benefits. In this context, model-based assessments need to assess the competition effects of various flexibility options for CES by taking the benefits and drawbacks into account. Additionally, the economic viability of CES systems needs to also be systematically determined by taking into account the corresponding learning rates. Thereby, the storage size reduction is indirectly related to the investment trends which again leads to a better viability even in terms of different flexibility options \cite{graditi2016technical}.

\subsection{Research scope}

In view of the reviewed papers and the mentioned research gaps, this paper aims to comprehensively demonstrate the operational and economic viability of CES systems for single-family households by taking competitive flexibility options into account. This paper answers the following two research questions:
\begin{itemize}
\item Flexibility competition effects: Which benefits and drawbacks are recognisable for CES systems at the household level, taking into account simultaneous flexibility options?
\item Storage market viability: When do the generated flexibility revenues of different scenarios exceed the assumed investments of CES systems, taking into account elaborated cost trends?
\end{itemize}

This paper is structured as follows: Section \ref{S:2} describes relevant storage concepts and application areas, and elaborates capacity cost trends. Section \ref{S:4} then introduces the optimization systematic, developes a cost function for CES and HES systems and outlines relevant indicators. Section \ref{S:5} presents the case studies of this paper. The optimization results are analyzed in section \ref{S:6}, which concludes with a discussion of the research results and limitations. Lastly, the results are summarized and future research is highlighted in Section \ref{S:8}.

\section{Storage conceptualization}
\label{S:2}

\subsection{Battery systems}
\label{S:2.1}
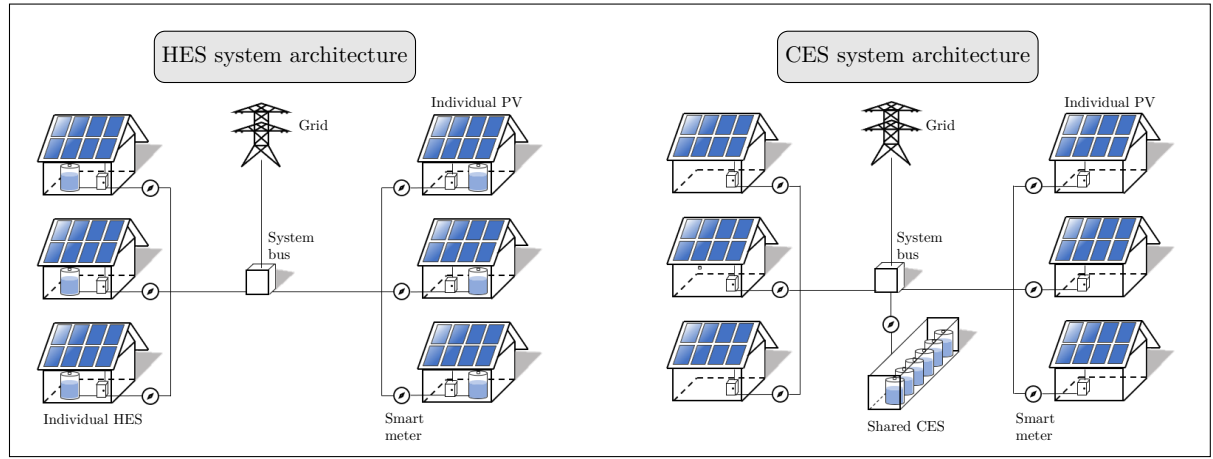
\begin{figure}[h]

\begin{tikzpicture}
\draw[black] (-8,-3.8) rectangle (7.8,2.2); 
\node[inner sep=0pt,scale=0.45] (distribution) at (0.5,-1.3) {\includegraphics[ width=2.2\textwidth]{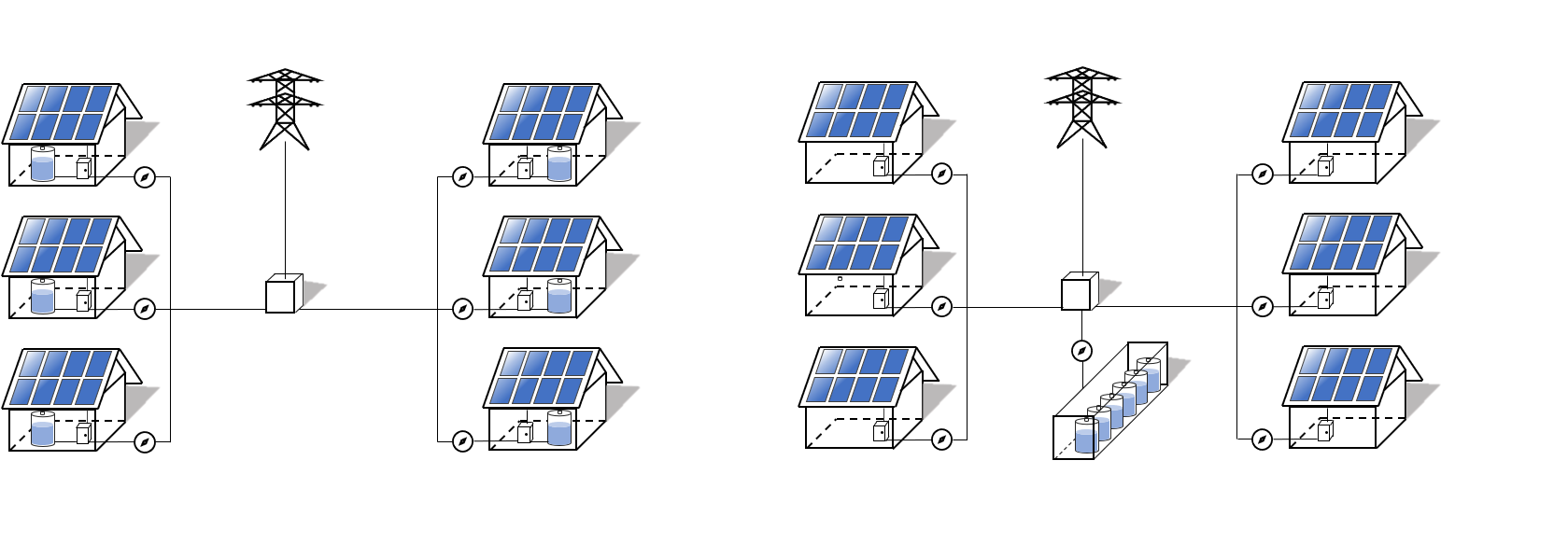}};
    
\tikzstyle{bbox} = [rectangle, rounded corners, minimum width=3cm, minimum height=0.8cm,text centered, draw=black, fill=black!10, scale=0.85]
\node (start) [bbox,right] at (-6.1,1.5) {HES system architecture};
\node (second) [bbox,right] at (2.1,1.5) {CES system architecture};

\node at (4.35,-1.) [left, scale=0.55,align =left] {System \\bus};
\node at (-3.92,-1.) [left, scale=0.55,align =left] {System \\bus};
\node at (-6.9,-3.3) [scale=0.55,align =left] {Individual HES};
\node at (3.8,-3.4) [scale=0.55,align =left] {Shared CES};
\node at (-1.85,0.9) [scale=0.55,align =left] {Individual PV};
\node at (-2.8,-3.4) [scale=0.55,align =left] {Smart \\meter};
\node at (-4.,0.6) [scale=0.55,align =left] {Grid};
\node at (4.25,0.6) [scale=0.55,align =left] {Grid};
\node at (6.48,0.9) [scale=0.55,align =left] {Individual PV};
\node at (5.5,-3.4) [scale=0.55,align =left] {Smart \\meter};
\end{tikzpicture}
\caption[Neighborhood sharing a CES]{Schematic representation of household electricity storage (HES) and community electricity storage (CES) systems (own representation based on \cite{MULLER2018492})} 
\label{foundation}

\end{figure}

Storage deployment at distribution grid level is expected to play an important role in the energy transition \cite{KOIRALA2018570,Braff.2016}. Within the BES domain, two concepts are emerging in the residential sector for single-family housing, namely household electricity storage (HES) and community electricity storage (CES) systems. Figure \ref{foundation} provides a schematic representation of both concepts.

A HES can be described as a stationary BES deployed in individual dwellings. Most HES systems serve self-consumption of locally-generated electricity, often deployed as a complementary solution to RE systems \cite{VANDERSTELT2018266, PARRA2017}. In this application, surplus electricity is not fed into the grid, but stored for consumption at a later point in time \cite{MULLER2018492}. A HES investment might be undertaken due to different motivations of end users, such as psychological, ecological or financial motives \cite{agnew2015, MULLER2018492}. 


A CES constitutes another promising approach. The definition of CES varies widely among scientific literature \cite{KOIRALA2018570}. In essence, two key aspects are mentioned regularly and are central for understanding the concept. First, a CES is defined as a storage system shared between members of a community who are typically, but not exclusively, located in spatial proximity \cite{BARBOUR2018}. Similarly to HES systems in individual dwellings, a CES gives energy communities greater control in managing RE collectively and locally. Emerging grassroot initiatives call for a fundamental shift of control, access and ownership structures in order to empower local communities \cite{Fairchild.2017}. Second, several scholars refer to CES as an intermediate link between end users in a community and the wider energy network \cite{KOIRALA2018570}. 

Recently, CES systems seem to receive more attention. Firstly, a shared operation of a BES levels out fluctuations of electricity demand and generation profiles of all end users \cite{Wawer.2018,Zeh.2016}. Secondly, scaling effects make CES systems economically more favorable \cite{PARRA2017aop,Wawer.2018}. Thirdly, beyond its primary purpose for MSC for the community, the CES could be utilized by multiple applications. An overview of selected CES projects in Germany is outlined in Table \ref{tab:pilots} \cite{Kress2014,Thomann2016a,IBCSOLARBlog2015,Jossen2015,Ullrich2016}. Notably, all projects were realized with Li-ion batteries. Apart from this, the overview highlights the diversity of CES configurations. \\

\begin{table}
\caption[Overview of selected CES projects in Germany]{Overview of selected CES projects in Germany (own representation based on \cite{Kress2014,Thomann2016a,IBCSOLARBlog2015,Jossen2015,Ullrich2016})}
\label{tab:pilots}
\centering
\begin{tabular}{@{}lllllll@{}} 
\hline
Operator        & Project         & Launch    & Storage                                                                          & Type   & Generator                                                                                 & User                                                                                                                                                                                       \\ 
\hline
KACO            & Weinsberg     & Nov. 2013 & 150 kWh                                                                          & Li-ion & \begin{tabular}[c]{@{}l@{}}145 kWp PV\\ 11 kWel CHP \end{tabular} & \begin{tabular}[c]{@{}l@{}}23 households\\  \end{tabular}  \\
MVV Energie     & Strombank     & Dec. 2014 & \begin{tabular}[c]{@{}l@{}}100 kWh\\100 kW\end{tabular} & Li-ion & \begin{tabular}[c]{@{}l@{}}PV\\ CHP \end{tabular}                 & \begin{tabular}[c]{@{}l@{}}14 households\\4 commercials \end{tabular}                                                                                             \\
IBC Solar       & Epplas        & Apr. 2015 & \begin{tabular}[c]{@{}l@{}}660 kWh\\ 70 kW\end{tabular}  & Li-ion & 287 kWp PV                                                                                & 13 households \\
Kraftwerke Haag & EEbat         & Oct. 2015 & \begin{tabular}[c]{@{}l@{}}250 kWh\\ 200 kW\end{tabular} & Li-ion & 300 kWp PV                                                                                & 20 households                                                                                                                                                                              \\
ENTEGA          & Solarsiedlung & Sep. 2016 & \begin{tabular}[c]{@{}l@{}}800 kWh\\250 kW\end{tabular} & Li-ion & PV & 82 households \\
\hline
\end{tabular}
\end{table}

\subsection{Application areas}
\label{S:2.2}

BES systems for households can be beneficial in numerous ways for different stakeholders depending on compatible applications \cite{phdmetz}. Table \ref{tab:applications} summarizes relevant application areas regarding practical implementation and commercial viability for Li-ion CES systems \cite{MULLER2018492}. Applications with a direct impact on the energy expenses of end user households (e.g. cost savings) are referred to as end user applications, while services with an impact on the municipal utility profits' (e.g. revenue streams) are referred to as operator applications.

\begin{table}[H]
\centering
\caption{Overview of applications for BES focusing on households (own representation based on \cite{MULLER2018492})}
\begin{tabular}{@{}p{6cm}p{10cm}@{}}
\toprule
End user applications       &         \\\cmidrule(lr){1-1}
Maximizing  self-consumption (MSC) &  Maximizing the consumption of on-site generated electricity by storing the surplus for subsequent use \\ \addlinespace[0.3cm]
Bill management (BM) & Storing electricity in periods with favorable prices, then consuming the energy during peak periods \\ \midrule
Operator applications       &         \\ \cmidrule(lr){1-1}
Spot market arbitrage (ARB)                 & Storing of electricity from the wholesale market in periods with favorable prices and selling the electricity during peak periods  \\ \addlinespace[0.3cm]
Frequency regulation (FR)    & Balancing out fluctuations of the frequency in a grid by charging or discharging the battery system   \\ \bottomrule
\end{tabular}
\label{tab:applications}
\end{table}

At the end user level, maximizing  self-consumption (MSC) and bill management (BM) can be used to add value. MSC is the most widely considered application \cite{phdmetz, PARRA2017}. The economic viability of MSC highly depends on the remuneration for exported electricity and retail prices of imported electricity \cite{phdmetz}. BM is also referred to as demand load shifting, time shifting or end consumer arbitrage in several studies and is only applicable if consumers face time-varying electricity tariffs or capacity related prices and need to reduce their peak \cite{kpmg, Stephan.2016}. 

On the grid level, spot market abitrage (ARB) (approx. \SI{17.7}{\text{\euro}\per\kW} p.a. \cite{zakeri2017economics}) as well as frequency regulations (FR) (approx. \SI{156}{\text{\euro}\per\kW} p.a. for primary, \SI{170}{\text{\euro}\per\kW} p.a. for secondary, and \SI{57}{\text{\euro}\per\kW} p.a. for tertiary control reserve \cite{Zeh2015}) are  applications that add value for the operator. ARB conceptually resembles BM. Contrary to BM, ARB applications are associated with a trading strategy to exploit the margin of price differentials of upper-level markets without the consumption of energy \cite{phdmetz}. CES systems can be charged with electricity from the wholesale market at times of low prices and discharged when the price is higher. FR, also referred to as reserve control, frequency control or more generally ancillary services, is one of the most attractive application \cite{PARRA2017, phdmetz}. 

Despite the promising approach of FR, the implementation is connected to high operational uncertainty. First of all, a CES needs to pass a pre-qualification procedure in order to be in a position to take part in the market. Furthermore, due to the stochastic nature of the reserve market, the participation might also lead to limitations to the level of residential self-consumption. This is especially true if positive frequency control is requested. Moreover, the design of the FR can vary greatly between different countries. For these reasons, the application MSC and ARB are in focus of the analysis. While the HES systems are only able to apply MSC, the CES systems are also able to apply ARB.

\subsection{Cost trends}
\label{S:3.1}

\begin{figure}[h!]
    \centering
    \includegraphics[width=\linewidth]{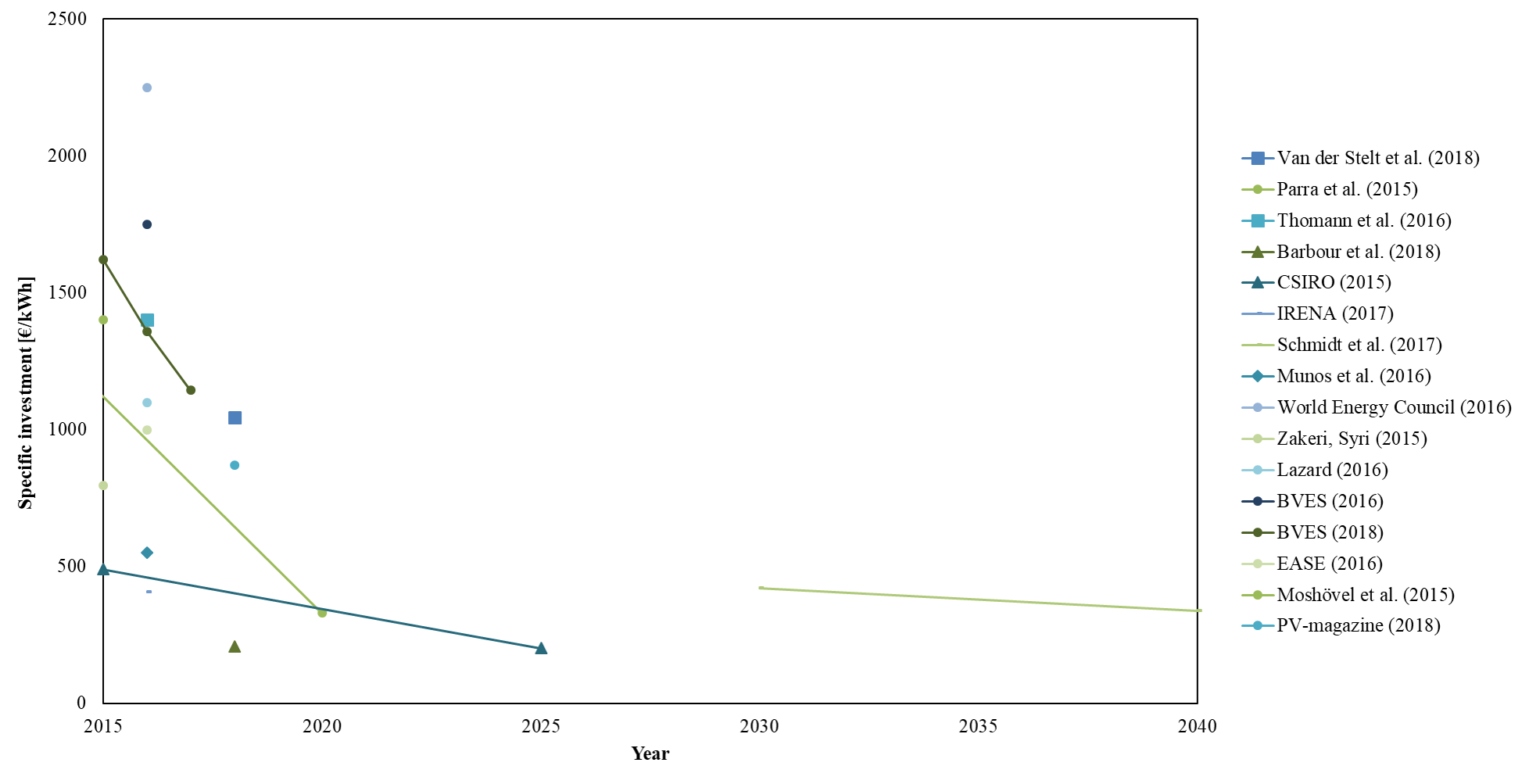}
    \caption{Projected price development of Li-ion batteries (own representation based on data points of \cite{VANDERSTELT2018266,PARRA2015,Thomann2016a,BARBOUR2018,Reedman2015,IRENA2017,MUNOZ2016,WEC2016,ZAKERI2015569,Lazard2016,schmidt2017future,BVES2016,BVES2017,EASE2016,moshovel2015,PVMAGAZINE2018})}
    \label{fig:future cost}
\end{figure}

Delivering a comprehensive and reliable picture of the cost development of energy storage systems is challenging. In the past 15 years, BES systems witnessed a development comparable to PV cells \cite{IRENA2017b}. The main technical drivers that led to cost reductions in the past are increased production scales, material improvements, more competitive supply chains, performance enhancements and benefits from operating experience \cite{IRENA2017b}. The technical drivers and the wide applicability of Li-ion batteries are expected to further reduce costs \cite{NREL2016}. Figure \ref{fig:future cost} depicts some projected price developments for stationary Li-ion batteries in \EUR{}/kWh\footnote{In order to display the prices in a unified currency, price information of different currencies have to be converted to \EUR{} by using the average exchange rate of the whole respective year in which the data are published. Furthermore, the assumptions is made that all prices are net prices.} from 2015 until 2040.


Nevertheless, a clear trend towards declining future prices is visible. In \cite{IRENA2017b}, the average battery prices are projected to reduce to about 60 \EUR{}/kWh capacity in the year 2030 in the best-case scenario \cite{IRENA2017c}. According to \cite{moshovel2015}, prices for Li-ion batteries will remain fairly stable after the year 2030 since market saturation will be reached.

The price ranges for inverters also differ strongly within the examined literature. In a report of CSIRO (2015) in which future trends for BES systems were investigated, the prices for inverters range from 514 \EUR{}/kWh in the best-case scenario up to 1410 \EUR{}/kWh in the worst-case scenario \cite{Reedman2015}. In comparison, \cite{IRENA2017c} quoted much lower prices for battery inverters ranging from about 105
\$/kW in the best case up to 290 \$/kW in the worst case for the year 2016 \cite{IRENA2017c}. These lower costs correspond to inverter cost information of \cite{moshovel2015} and \cite{OKOINSTITUT2018}.

In the assessment of this paper, the economic performance of HES and CES is compared for three reference years 2015, 2025 and 2035. While an average price scenario is chosen for storage systems and inverters, the proportionate cost decline from 2015 to 2025 is significantly higher compared to the cost decline from 2025 to 2035. The average case is assumed as follows: 2015 - Storage 500 \EUR{}/kW, Inverter 200 \EUR{}/kW; 2025 - Storage 200 \EUR{}/kW, Inverter 100 \EUR{}/kW; 2035 - Storage  150 \EUR{}/kW, Inverter 70 \EUR{}/kW.

\section{Model approach}
\label{S:4}

\subsection{Optimization model}
\label{SS:4-1}

Given the complexity of CES systems, decision-makers increasingly require adequate assessment methods. To answer specific questions of relevance at the municipal level, the operation-oriented optimization model IRPopt (Integrated Resource Planning and Optimization) has been developed to sufficiently map community energy system complexities \cite{SCHELLER2018,kuehne2019}. A proven graph-based energy system approach allows the accurate modeling of deployment systems by considering different energy carriers and technical processes. In addition, a graph-based commercial association approach enables the integration of actor-oriented coordination. This is achieved by the explicit modeling of municipal market actors on one layer and technology processes on another layer as well as resource flow interrelations and commercial agreements mechanism among and between the different layers. IRPopt represents a bottom-up techno-economic optimization model, implemented in GAMS/CPLEX, for solving mixed-integer problems (MIP). The major objective is to maximize revenues from different actor perspectives. This leads to optimally dispatched energy flows between the technical components for individual households but also the whole interrelated community. In this context, the objective function exhibits a novel formal interface between the supply and demand side which merges the modeled technical and economical aspects. For application, the time-steps of the scheduling horizon - one year of 35040 quarter-hourly time steps - are subdivided to enable a recursive dynamical optimization. The mathematical formulation of the model design as well as all relevant model equations have been published elsewhere \cite{SCHELLER2018,scheller2019entity,phdscheller}. Furthermore, all relevant equations of IRPopt are outlined in the Supplementary Material.

In the framework of this paper, the model works with an actor-related two-step optimization approach. The model firstly optimizes the energy flows from an aggregated customer perspective, determining the residual energy demand and excess energy supply with all components the customers have access to. The specified load demand of the households needs to be covered with own generation or grid energy imports. The demand side flexibility options can be applied to reduce the costs. With respect to the first optimization step (prosumer optimization), especially the tariff scheme as well as the variable costs of decentral energy technologies are decisive. In the subsequent step, the model optimizes all other energy from the utilities’ perspective, considering the total residual energy demand and supply. A community related energy deficit might be balanced by storage systems, generation plant activities and spot market trading. Excess energy is sold to the spot market. With respect to the second optimization step (operator optimization), especially the market prices of the central energy systems are decisive. 


\subsection{Cost function}
\label{S:3.2}
For a systematic comparison of CES and HES systems, it is assumed that both technologies have the same basic structural design. According to \cite{Thomann2016a} and \cite{zhu2011case}, the core of a CES is the battery storage consisting of several battery cells. Furthermore, a battery management system to control the battery and a power conversion unit is needed. Necessary components additionally comprise a gaseous fire suppression, air conditioning, a firewall and smart meters. In this context, the general cost function for HES ($Cost_{tot}^{HES}$) and CES ($Cost_{tot}^{CES}$) systems is stated in Equation \ref{eq:cost_BES}.

\begin{equation}\label{eq:cost_BES}
        Cost_{tot}^{HES|CES} = Cost_{sm}^{HES|CES} + Cost_{inv}^{HES|CES}
\end{equation}

The first part of the cost function to be developed are the costs for the Li-ion cells ($Cost_{sm}^{HES|CES}$) as outlined in Equation \ref{eq:cost_sm2}. Since the data basis was characterized by high opacity, a meaningful distinction between HES and CES system cost per kWh is not possible. Nevertheless, authors agree that larger systems in terms of their capacity are relatively cheaper than smaller ones due to economies of scale \cite{ISEA2018, PARRA2015, Sieg2015, IWES2014}. Considering this, \cite{PARRA2015} applied a power factor of $0.7$ for BES systems which have a larger capacity ($C_{nom}$) than a certain threshold ($C_{l}$). Relative storage costs in \EUR{}/kWh are given by $cost_{sm}$. \cite{PARRA2015} assume 100 kWh for $C_{l}$ . Since the BES capacities are smaller in this paper, it is assumed that economies of scale are already realized for BES systems larger than 10 kWh. To compensate for the decreased capacity threshold $C_{l}$, the power factor is modified from $0.7$ from \cite{PARRA2015} to $0.9$ in this paper.

\begin{equation}\label{eq:cost_sm2}
Cost_{sm}^{HES|CES}=\left[ \left( \frac{ C_{nom} }{ C_{l} } \right)^{ 0.9 } \right]\cdot C_{l} \cdot cost_{sm}
\end{equation}

The second cost component  ($Cost_{inv}^{HES|CES}$) are bi-directional inverters for HES and CES systems as outlined in Equation \ref{eq:cost_inv}. Whilst for HES systems each one must be equipped with an inverter, a CES system can be used by the community with one inverter only \cite{GRIESE2016}. In terms of the investments, \cite{BARBOUR2018} state that the inverter has to be sized in kW to match the maximum charge and discharge rates of battery storages. Hence, as a reasonable and typical estimate for stationary Li-ion batteries they determined a value of 0.5 of the maximum state of charge (SOC) of the battery. Apart from the inverter sizing, they assumed that economies of scale may be realized for inverters having a capacity larger than 10 kW by the power of 0.7
 \cite{BARBOUR2018}. $C_{k}$ represents the threshold for economies of scale in kW and $cost_{inv}$ represents the relative inverter costs in \EUR{}/kW.

\begin{equation}\label{eq:cost_inv}
Cost_{inv}^{HES|CES}=\left[ \left( \frac{ 0.5\cdot SOC_{max} }{ C_{k} } \right)^{ 0.7 } \right]\cdot C_{k} \cdot cost_{inv}
\end{equation}

Additionally, in this paper the cost function also consists of operation and maintenance costs ($Cost_{om}^{CES}$). Little and ambiguous information is available according to \cite{Reedman2015}. O\&M costs are highly dependent on the operation of the BES but stationary Li-ion batteries also exhibit very low O\&M costs \cite{OKOINSTITUT2018}. Because larger storage facilities can be expected to require higher O\&M efforts \cite{MUNOZ2016}, HES systems are assumed to be free of O\&M costs as stated in \cite{BECK2017604, Lazard2016}, whilst annual O\&M costs for CES are assumed to be 1.5 \% of the investment of the cells and the inverter \cite{IRENA2017b,WEC2016}.

\subsection{Assessment indicators}
\label{sec:keyindicators}
To carry out a techno-economic assessment, a set of economic indicators (annual financial benefit, equivalent annual cost, equivalent annual value) and performance indicators (self-consumption ratio, the self-sufficiency ratio as well as the utilization ratio) are used. 

\subsubsection{Annual financial benefit}
The optimization approach of IRPopt determines the financial benefits for each involved actor of the dispatched energy flows. Since the optimization period is set to one year, the values represent an annual quantity and are referred to as annual financial benefit (AFB). In the HES scenario, AFB accrues only to the end users (AFB\textsubscript{eu}) and are equivalent to the energy cost savings of each prosumer. In the CES scenario, AFB can accrue not only to the end users but also to the operator (AFB\textsubscript{op}) of the CES.

\subsubsection{Equivalent annual cost}
In order to evaluate the total benefits of a system, the corresponding cost of each system over the same period must be determined. The equivalent annual cost (EAC) of each HES system and CES system share is expressed by Equation \ref{eq:EAChes}. In this context, the capital recovery factor is used to annuitize the storage investment. The composition of the total costs of the HES system ($Cost_{tot}^{HES}$) and of the CES system ($Cost_{tot}^{CES}$) based on the storage size is presented in Section \ref{S:3.2}. This method has also been used by \cite{KLINGLER20181064,BARBOUR2018,QUOILIN201658}. 

\begin{equation}
EAC^{HES|CES}= \left( \frac{(1+i)^n \cdot i}{(1+i)^n-1}\right)\cdot Cost_{tot}^{HES|CES} + Cost_{om}^{CES} \label{eq:EAChes}\\
\end{equation}  

\subsubsection{Equivalent annual value}
The total equivalent annual value (EAV) is calculated by subtracting the EAC from the AFB. If the EAV is positive, it is beneficial to install a BES system from an economic perspective. This calculation constitutes a cost-benefit-analysis on an annual basis as cost savings and revenues arising from the operation of the BES are compared to the annual cost for the battery. A similar approach is used by \cite{BARBOUR2018}.



\subsubsection{Self-consumption ratio}
The CES is primarily (and the HES exclusively) used to maximize the self-consumption of self-generated electricity of the households in order to reduce the operational energy costs of the end user. Thus, a main Figure of interest is the volume of SC, which can be expressed in absolute or relative terms. For a relative assessment, the SCR is commonly used in studies such as \cite{TERLOUW2019356,KLINGLER20181064,VANDERSTELT2018266}. In this paper, the SCR\textsubscript{tot} is defined as the ratio between self-consumed PV energy and the total annual electricity produced by the PV array. The PV energy can be used directly for the electric load  $E_{EL}^{PV}$, the $E_{HP}^{PV}$ or charging the battery $E_{EL}^{
CES|HES}$ for subsequent use. The SCR\textsubscript{tot} is described mathematically by Equation \ref{eq:SCR} below.

\begin{equation}\label{eq:SCR}
SCR_{tot}= \frac{E_{EL}^{PV}+E_{HP}^{PV}+E_{HES|CES}^{PV}}{E_{PV}} 
\end{equation}

To study how much each component contributes to the SCR\textsubscript{tot}, the SCR\textsubscript{EL}, SCR\textsubscript{HP} and SCR\textsubscript{CES} can also be calculated individually as presented in Equation \ref{eq:SCCES}. 

\begin{equation}\label{eq:SCCES}
SCR_{EL}= \frac{E_{EL}^{PV}}{E_{PV}} \quad;\quad SCR_{HP}= \frac{E_{HP}^{PV}}{E_{PV}}\quad;\quad
SCR_{HES|CES}= \frac{E_{HES|CES}^{PV}}{E_{PV}}
\end{equation}

\subsubsection{Self-sufficiency ratio}
The self-sufficiency ratio (SSR) is a metric used to express the energy autonomy of a household. In this paper, the SSR is defined as the share of electricity demand met by local generation. Thus, the SSR can be calculated according to Equation \ref{eq:SSR}. The energy consumed by the grid is represented as E\textsubscript{GC}, the total electricity demand is given by the sum of E\textsubscript{EL} and E\textsubscript{
HP}.

\begin{equation}\label{eq:SSR}
SSR= 1-\frac{E_{GC}}{E_{EL}+E_{HP}} 
\end{equation}

\subsubsection{Utilization ratio}

The storage utilization ratio (UR) gives an indication how well the available storage capacity is used. In this paper, the UR is defined as the number of time steps in which the storage is charged (T\textsubscript{charged}) divided by the total time steps (T\textsubscript{all}) as outlined in Equation \ref{eq:UR}.

\begin{equation}\label{eq:UR}
UR= \frac{T_{charged}}{T_{all}} 
\end{equation}

\section{Case studies}
\label{S:5}

\subsection{Model input}\label{subsec:modelinput}
In order to improve transparency and traceability of the results, the relevant input data of the scenarios is outlined in the following.

\subsubsection{Market parameters}
\label{SS:marketparameters}
\begin{figure}[h]
\centering
\includegraphics[width=\linewidth, height=\textheight,keepaspectratio]{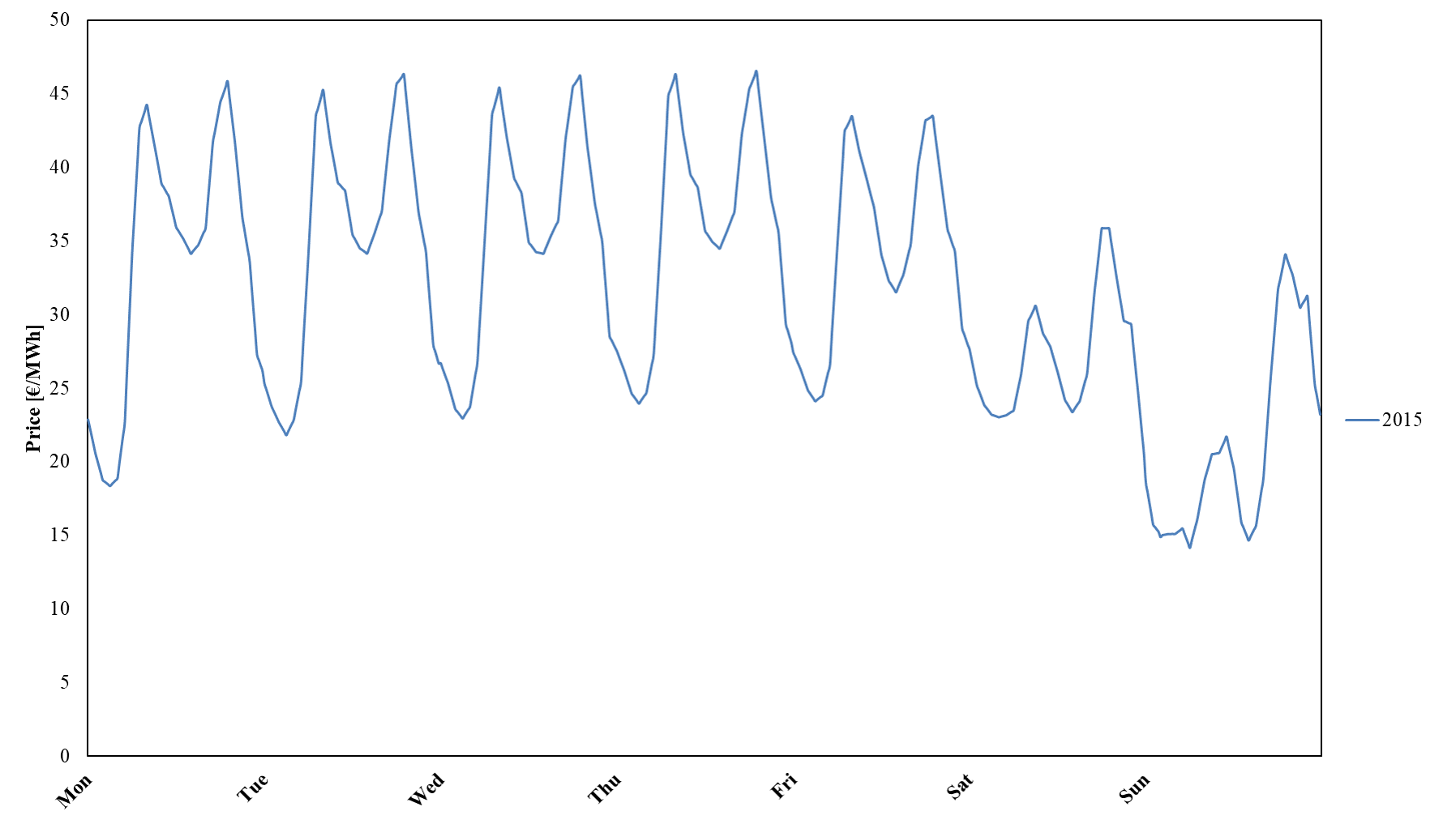}
\caption{Average week of EPEX spot market prices for the year 2015 (own representation based on EPEX data)}
\label{fig_spot}
\end{figure}

The quarter-hourly data sets are based on historical market data of 2015 of Germany. An overview of average spot market price characteristics is given in Figure \ref{fig_spot}. For the annuity an interest rate of $4\%$ per year is applied. For the value added tax $19\%$ is applied.

Furthermore, the scenarios include a flat tariff scheme for grid consumption as well as feed-in remunerations. The tariffs are composed of competitive pricing elements as well as statutory fees and levies as elaborated for Germany in \citep{Scheller.2018legal}. However, charging and discharging the CES system for self-consumption is assumed to be exempted from both the EEG (German Renewable Energy Sources Act) surcharge and electricity tax according to a recommendation of the German Energy Storage Association. A CES would not be utilized at all in terms of the existing legal specifications of Germany due to the double burden regarding grid related levies \cite{Scheller.2018legal}. Given the mean spot market price on the EPEX SPOT in 2015 of 3.16 Ct/kWh\textsubscript{el} (el =electrical), the remaining 3.96 Ct/kWh\textsubscript{el} of the sales cost component is assumed to be the utility margin. While the grid and regulatory components were kept constant in absolute numbers, the sales component was derived from the projected mean spot price and the fixed margin. Thus, the average cost of grid electricity amounts to 29.16 ct/kWh for a residential household. In this context, legally prescribed components currently account for around 55$\:$\% and grid fees for around 25$\:$\% of the electricity price. A similar procedure was applied to the cost components of the gas tariff. The initial price is 6.6 Ct/kWh\textsubscript{f} (f = fuel) with grid and regulatory fees and levies of 2.4 Ct/kWh\textsubscript{f} and a margin of 1.1 Ct/kWh\textsubscript{f}. For electricity fed into the grid from a residential PV system, the prosumer receives 0.124 \EUR{}/kWh\textsubscript{el} as a remuneration. The feed-in remuneration has been projected with the help of \citep{AgoraEnergiewende.2016}.

\subsubsection{Household parameters} 
\label{SS:householdparameters}
The community consists of six individual households. The demand of each household is represented by an electrical demand (el) profile based on the highly-resolved profile set of \citep{online:tjaden}. The average energy consumption per year varies between 2900 kWh\textsubscript{el} and 4500 kWh\textsubscript{el}. Each el profile is divided into a shiftable and non-shiftable share as outlined in the source \citep{Scheller.2018b}. This becomes relevant in scenario 3 and 4 as DR is applied. Since DR is associated with a level of discomfort, it is assumed that a maximum of $40\%$ of the electrical load can be shifted for 4\;\!h (DR\textsubscript{max}). A bottom-up explanation is given by \citep{Scheller.2018b}.

The heating demand of each household is represented by varying thermal load (th) profiles which were deduced on the basis of the Hellwig methodology \cite{phdhellwig}.
The annual average heat demand of each household ranges between 15,300 kWh\textsubscript{th} and 18,300 kWh\textsubscript{th}.

Moreover, each prosumer might be equipped with conversion processes and is connected to the storage process via the electrical grid (EG). Technology specification for the photovoltaic system (PV), the heat pump system (HP), the community electrical storage system (CES), and the natural gas boiler system (NGB) are outlined in Table \ref{technologies}. The given data is broken down to one single customer. The HP applied in this paper is a ground-source HP. In the scenarios without a HP, a natural gas boiler (NGB) is present to cover the thermal demand. The indices el and th are again related to the electrical or thermal aspects.

\begin{table}[H]
\centering
\caption{Technical characteristics of analyzed technologies (based on \cite{BVES2016,fraunumsicht,IWES2014})}
\begin{tabular}{@{} lcccccc@{}}
\toprule

Characteristics & PV & HP &   NGB  \\ \midrule
Area & 0-30 m\textsuperscript{2} &  &  \\
Power &  & 9 kW\textsubscript{th}    & 9 kW\textsubscript{th}   &  \\
Efficiency  & 18 \%\textsubscript{el} & 95 \%\textsubscript{el}  & 95 \%\textsubscript{th} &  \\
Coefficient of performance (COP)  &  & 3.9 &     \\
 \bottomrule
\end{tabular}
\label{technologies}
\end{table}

Li-ion batteries are considered for the BES systems. The charging and discharging efficiency of the BES system is assumed to be 95\% \cite{BVES2016,fraunumsicht,IWES2014}. The expected self-discharge of Li-ion batteries is very low (2\% to 5\% per month) and was set to 0.001\% per each quarter-hourly timestep, equaling just below 3\% per month \cite{BVES2016,fraunumsicht}. Furthermore, the charge and discharge capacities are scaled according to the BES sizes. Table \ref{escharacterisitcs} provides an overview of the relevant technical characteristics of BES systems.

\begin{table}[H]
\centering
\caption{BES characteristics (based on \cite{BVES2016,fraunumsicht,IWES2014})}
\begin{tabular}{@{} lll@{}}
\toprule
Characteristics & Value& Unit   \\ \midrule
(Dis)charging efficiency & 95 &\%   \\
Self-discharging rate  & 0.001 & $\%/1/4h$  \\
Technical lifetime  & 20 & a  \\
 \bottomrule
\end{tabular}
\label{escharacterisitcs}
\end{table}

\subsection{Scenario overview}
The system design considered regarding the scenarios applied for each of the HES and CES cases is illustrated in Figure \ref{foundation}. The presented configuration represents also the baseline scenario (scenario 1). In both cases a community or rather neighborhood comprising of six households has been modelled. Thereby, IRPopt allows to model each household on his own. Furthermore, a municipal utility which also acts as CES operator is providing the community. Each household has an electricity demand which is represented by a heterogeneous load profile as outlined in Section \ref{SS:householdparameters}. The load can be covered by grid consumption for the given tariffs as outlined in Section \ref{SS:marketparameters} or self-consumption. Excess PV energy can be stored temporarily or fed into the grid. The technology portfolio for self-consumption among the households is identical in each scenario. In the HES case, HES are installed within the premises of each household. In the CES case, a central storage system is installed in spatial proximity and households get an equal share of the CES. The CES is managed by an operator and the end users of the community are connected to the CES via the public grid. The aggregated capacity of all HES systems spread over the six households equals the total CES capacity.

Each household of the HES cases exclusively uses its HES to store excess PV energy for subsequent use which goes in line with a strategy of MSC. In contrast, in the CES cases, the CES can perform several applications with a positive impact for the households and the operator. First, end users can utilize the CES system for MSC. Second, an operator can generate revenues with upper-stream applications in time steps when the system is underutilized. In terms of this paper, ARB can be applied. Since other flexibility options could reduce the market viability of a CES, in certain scenarios participating households also have the possibility to apply different flexibility options.

In total, five scenarios are modelled and analyzed as outlined in Table \ref{cases}. IRPopt accordingly optimize the energy flows of the configurations. Scenario 1 is considered, as already stated, as the baseline scenario serving as a benchmark. Scenarios 2-4 serve to evaluate the impact of flexibility options on the techno-economic performance of CES systems, while scenario 5 is used to study the influence of a crucial CES system design parameter. In general, the different scenarios are of the same design. However, the technology portfolio of the six household prosumers is changing between the scenarios 1-5.

In scenario 1, the community is structured according to the descriptions above. Each household possess a PV unit and a connection to the HES or CES system. Excess PV energy can be stored temporarily or fed into the grid. In scenario 2, a HP is added to the prosumer portfolio of each community member for electrical heat generation. In this scenario, excess PV electricity can be used for heat conversion via the on-site HP rather than exporting it to the grid or charging the CES system. In scenario 3, DR  is added to provide additional flexibility and in scenario 4 both flexibility options are applied nest to the CES system. Finally, in scenario 5 the power-to-energy (P/E)-ratio of the BES systems is changed. The P/E ratio constitutes a measure of the charge/discharge power compared to the energy storage capacity \cite{SEFEP}. Thus, in scenarios 1-4 the P/E ratio is set to 1 meaning that the charging and discharging powers are set according to the respective storage capacities. A insight into the exact mathematical modelling of the scenarios is given in the Supplementary Material.

\begin{table}[H]
\centering
\caption{Configuration of scenarios applied}
\label{cases}
\begin{tabular}{@{}lcccccc@{}}
\toprule
          &  & \multicolumn{2}{c}{Flexibility option}                                   &  & \multicolumn{2}{c}{P/E ratio [kW/kWh]}                               
          \\ \cmidrule{3-4} \cmidrule{6-7} 
          &  & \multicolumn{1}{c}{HP} & \multicolumn{1}{c}{DR\textsubscript{max}}  &  &
          \multicolumn{1}{c}{1} & \multicolumn{1}{c}{0.5} 
          \\ \midrule
Scenario 1: Base Case &  
& \ding{55} &  \ding{55} &  &  \ding{51} &  \ding{55}  
\\\hdashline
Scenario 2: HP Case &  &  \ding{51} & \ding{55}&  &\ding{51}   &  \ding{55} 
\\ \hdashline
Scenario 3: DR Case &   &  \ding{55} &   \ding{51}&  & \ding{51} & \ding{55} 
\\\hdashline
Scenario 4: HP + DR Case &   &  \ding{51} &   \ding{51}&  & \ding{51} & \ding{55} 
\\\hdashline
Scenario 5: Power-to-Energy Case &   &  \ding{55} &   \ding{55}&  & \ding{55} & \ding{51} 
\\  \bottomrule
\end{tabular}
\end{table}

According to \cite{HOPPMANN20141101} the financial return is highly sensitive to different PV and HES or CES combinations. Thus, a sensitivity analysis is conducted. The HES and CES share for each household is varied from 0 to 10 kWh with steps of 2.5 kWh. The different battery capacities are combined with varying PV sizes, ranging from 0  to 30 m\textsuperscript{2} in steps of 7.5 m\textsuperscript{2}. The five different PV sizes in combination with five different BES capacities result in 25 sensitivities in total for each of the HES and CES scenario. 

\section{Optimization results and discussion}
\label{S:6}

\subsection{Techno-economic assessment}
The annual optimization results of the different scenarios are initially analyzed with respect to the techno-economic indicators AFB, EAC, EAV, and UR. While Sections \ref{sec:6-1} and \ref{sec:performanceanalysis} consider the results of scenario 1 in detail, Section \ref{sec:6-3} compares the results to scenarios 2-5. 

\subsubsection{Flexibility analysis}
\label{sec:6-1}
The results of the AFB demonstrate the financial benefits obtained when operating a BES. An overview is given in Table \ref{tab:baselineAFB}. In the case of HES, the value corresponds to the annual energy saving for each household enabled by the operation of the HES in order to offset grid consumption. Thereby, only the benefits directly related to the storage application are considered. This relates to savings of the retail prices but also regulatory levies and grid fees. Direct self-consumption benefits of self-generated energy are not included. In the case of the CES, the AFB represents the same yearly energy saving of each community member including the additional revenues attained by the operator in CES idle-times due to operator applications as well as losses due to less grid consumption by the community members.

\begin{table}[H]

\centering
\caption{ Comparison of aggregated form of HES and CES systems (operator and prosumer results) based on AFB (the CES size[kWh] is given per household)}
\begin{tabular}{@{}llrrrrr@{}}
\toprule
                      & \multicolumn{6}{c}{\qquad HES [$kWh$]}      \\ \cmidrule{3-7}
\multirow{9}{*}{\rotatebox[origin=c]{90}{~PV [$m^2$]}} &    & \multicolumn{1}{c}{0} & \multicolumn{1}{c}{2.5} & \multicolumn{1}{c}{5} & \multicolumn{1}{c}{7.5} & \multicolumn{1}{c}{10} \\ \cmidrule{2-7}
& 0  & -   &   \multicolumn{1}{c}{-}    & \multicolumn{1}{c}{-}    &  \multicolumn{1}{c}{-}   &   \multicolumn{1}{c}{-}  \\\cmidrule{2-7}
& 7.5 & -   &46 \EUR{}    &  52 \EUR{} &  52 \EUR{}   &  52 \EUR{}  \\\cmidrule{2-7}
& 15   &  -  &86 \EUR{}     & 119 \EUR{}  &  130 \EUR{}   & 133 \EUR{}   \\\cmidrule{2-7}
& 22.5 & -    & 103 \EUR{}     & 151 \EUR{}  &  170 \EUR{}   &   177 \EUR{} \\\cmidrule{2-7}
& 30  & -    & 111 \EUR{}    & 169 \EUR{}  & 192 \EUR{}    & 201 \EUR{}  \\ \bottomrule
\end{tabular}
\quad
\begin{tabular}{@{}llrrrrr@{}}
\toprule
                      & \multicolumn{6}{c}{\qquad CES [$kWh$]}      \\ \cmidrule{3-7}
\multirow{9}{*}{\rotatebox[origin=c]{90}{~PV [$m^2$]}} &    & \multicolumn{1}{c}{0} & \multicolumn{1}{c}{2.5} & \multicolumn{1}{c}{5} & \multicolumn{1}{c}{7.5} & \multicolumn{1}{c}{10} \\ \cmidrule{2-7}
& 0  & -   &   28 \EUR{}   & 55 \EUR{}   &   83 \EUR{}   &   110 \EUR{}  \\\cmidrule{2-7}
& 7.5 & -    &61 \EUR{}    &  93 \EUR{} &  121 \EUR{}   &  149 \EUR{}  \\\cmidrule{2-7}
& 15   &  -   &90 \EUR{}     & 141 \EUR{}  &  177 \EUR{}   & 207 \EUR{}   \\\cmidrule{2-7}
& 22.5 & -   & 102 \EUR{}     & 164 \EUR{}  &  206 \EUR{}   &   238 \EUR{} \\\cmidrule{2-7}
& 30  & -   & 108 \EUR{}    & 177 \EUR{}  & 222 \EUR{}    & 256 \EUR{}  \\ \bottomrule
\end{tabular}
\label{tab:baselineAFB}
\end{table}


For almost all system combinations the operation of one CES system attains higher financial benefits per year than operating multiple HES systems. Considering only the sensitivities in which a PV and a CES system is present (16 out of 25 combinations), the results demonstrate that the AFB can be increased by 29\% on average (avg. HES: 122 \EUR{} versus avg. CES: 157 \EUR{}). This improvement can be attributed to the outlined multi-application utilization of the CES system as assumed in this paper. These findings support the idea of previous research \cite{LOMBARDI2017, MULLER2018492, HE20111575,PARRA20191343,braeuer2019battery} that a multi-use of BES systems can improve the economic viability. Different scholars analyzed the effect of combining different applications  for residential batteries and found that a combination of MSC with BM increases the profitability of CES systems by 29\% compared to a CES used for MSC only \cite{PARRA20191343}. Despite different research scopes and considered applications, \cite{PARRA20191343} report similar numerical evidence for the positive effect of combining applications on profitability.

In this matter, the sensitivities with larger PV systems (22.5 m\textsuperscript{2} and 30 m\textsuperscript{2}) in combination with the smallest battery sensitivity (2.5 kWh) lead to worse results regarding AFB when using one CES instead of six HES systems (HES: 111\EUR{} versus CES: 108\EUR{}). To explain these results, the AFB obtained by both actor groups (e.g. end users and operator) from the CES, which is displayed in Table \ref{tab:baselineAFB} in aggregated form, must be considered separately. In these cases, the AFB from the operator AFB\textsubscript{op} is negative for the CES share compared to the HES case. This can be attributed to the fact that the utility or rather operator suffers significant losses in retail as end users cover their demand. Additionally, the comparably large PV-modules in contrast to small storage capacities lead to a higher loading of the CES with PV electricity. Therefore, the utility cannot compensate the missed sales by using the CES for its own purposes. 

A comparison of the EAC of HES and CES systems has been conducted with the help of the developed cost formula. Table \ref{tab:baselineEAC} shows that for each storage capacity the CES is more favourable than an individual HES. This can be explained by the economies of scale which accounts for price reductions with increasing sizes. The EAC can be reduced on average by around 14\% with a CES. 

\begin{table}[H]

\centering
\caption{Comparison of aggregated form of HES and CES systems (operator and prosumer results) based on EAC (the CES size[kWh] is given per household)}
\begin{tabular}{@{}llrrrrr@{}}
\toprule
                      & \multicolumn{6}{c}{\qquad HES [$kWh$]}      \\ \cmidrule{3-7}
\multirow{9}{*}{\rotatebox[origin=c]{90}{~PV [$m^2$]}} &    & \multicolumn{1}{c}{0} & \multicolumn{1}{c}{2.5} & \multicolumn{1}{c}{5} & \multicolumn{1}{c}{7.5} & \multicolumn{1}{c}{10} \\ \cmidrule{2-7}
& 0  & -   &   123 \EUR{}   & 246 \EUR{}   &   368 \EUR{}   &   491 \EUR{}  \\\cmidrule{2-7}
 & 7.5 & -    &123 \EUR{}    &  246 \EUR{} &  368 \EUR{}   &  491 \EUR{}  \\\cmidrule{2-7}
& 15   &  -   &123 \EUR{}     & 246 \EUR{}  &  368 \EUR{}   & 491 \EUR{}   \\\cmidrule{2-7}
& 22.5 & -    & 123 \EUR{}     & 246 \EUR{}  &  368 \EUR{}   &   491 \EUR{} \\\cmidrule{2-7}
& 30  & -   & 123 \EUR{}    & 246 \EUR{}  & 368 \EUR{}    & 491 \EUR{}  \\ \bottomrule
\end{tabular}
\quad
\begin{tabular}{@{}llrrrrr@{}}
\toprule
                      & \multicolumn{6}{c}{\qquad CES [$kWh$]}      \\ \cmidrule{3-7}
\multirow{9}{*}{\rotatebox[origin=c]{90}{~PV [$m^2$]}} &    & \multicolumn{1}{c}{0} & \multicolumn{1}{c}{2.5} & \multicolumn{1}{c}{5} & \multicolumn{1}{c}{7.5} & \multicolumn{1}{c}{10} \\ \cmidrule{2-7}
& 0  & -   &   121 \EUR{}   & 223 \EUR{}   &   315 \EUR{}   &   402 \EUR{}  \\\cmidrule{2-7}
& 7.5 & -   &   121 \EUR{}   & 223 \EUR{}   &   315 \EUR{}   &   402 \EUR{}  \\\cmidrule{2-7}
& 15   & -   &   121 \EUR{}   & 223 \EUR{}   &   315 \EUR{}   &   402 \EUR{}  \\\cmidrule{2-7}
& 22.5   & -   &   121 \EUR{}   & 223 \EUR{}   &   315 \EUR{}   &   402 \EUR{}  \\\cmidrule{2-7}
& 30  & -   &   121 \EUR{}   & 223 \EUR{}   &   315 \EUR{}   &   402 \EUR{}   \\ \bottomrule
\end{tabular}
\label{tab:baselineEAC}
\end{table}


The EAV can be subsequently calculated with the help of the yearly benefits presented in Table \ref{tab:baselineAFB} minus the yearly costs in Table \ref{tab:baselineEAC}. The results provided in Table \ref{tab:baselineEAV} indicate that neither HES nor CES systems are profitable in any combination in the year 2015. 

\begin{table}[H]

\centering
\caption{ Comparison of aggregated form of HES and CES systems (operator and prosumer results) based on EAV (the CES size[kWh] is given per household)}
\begin{tabular}{@{}llrrrrr@{}}
\toprule
                      & \multicolumn{6}{c}{\qquad HES [$kWh$]}      \\ \cmidrule{3-7}
\multirow{9}{*}{\rotatebox[origin=c]{90}{~PV [$m^2$]}} &    & \multicolumn{1}{c}{0} & \multicolumn{1}{c}{2.5} & \multicolumn{1}{c}{5} & \multicolumn{1}{c}{7.5} & \multicolumn{1}{c}{10} \\ \cmidrule{2-7}
& 0  & -   &   -123 \EUR{}   & -246 \EUR{}   &   -368 \EUR{}   &   -491 \EUR{}  \\\cmidrule{2-7}
 & 7.5 & -    &-77 \EUR{}    &  -194 \EUR{} &  -316 \EUR{}   &  -439 \EUR{}  \\\cmidrule{2-7}
& 15   &  -   &-37 \EUR{}     & -126 \EUR{}  &  -238 \EUR{}   & -358 \EUR{}   \\\cmidrule{2-7}
& 22.5 & -    & -20 \EUR{}     & -94 \EUR{}  &  -198 \EUR{}   &   -314 \EUR{} \\\cmidrule{2-7}
& 30  & -   & -12 \EUR{}    & -77 \EUR{}  & -176 \EUR{}    & -291 \EUR{}  \\ \bottomrule
\end{tabular}
\quad
\begin{tabular}{@{}llrrrrr@{}}
\toprule
                      & \multicolumn{6}{c}{\qquad CES [$kWh$]}      \\ \cmidrule{3-7}
\multirow{9}{*}{\rotatebox[origin=c]{90}{~PV [$m^2$]}} &    & \multicolumn{1}{c}{0} & \multicolumn{1}{c}{2.5} & \multicolumn{1}{c}{5} & \multicolumn{1}{c}{7.5} & \multicolumn{1}{c}{10} \\ \cmidrule{2-7}
& 0  & -   &   -93 \EUR{}   & -167 \EUR{}   &   -232 \EUR{}   &   -292 \EUR{}  \\\cmidrule{2-7}
& 7.5 & -   &   -60 \EUR{}   & -130 \EUR{}   &   -194 \EUR{}   &   -254 \EUR{}  \\\cmidrule{2-7}
& 15   & -   &   -31 \EUR{}   & -81 \EUR{}   &   -137 \EUR{}   &   -195 \EUR{}  \\\cmidrule{2-7}
& 22.5   & -   &  -19 \EUR{}   & -58 \EUR{}   &   -109 \EUR{}   &   -164 \EUR{}  \\\cmidrule{2-7}
& 30  & -   &   -13 \EUR{}   & -45 \EUR{}   &   -93 \EUR{}   &   -147 \EUR{}   \\ \bottomrule
\end{tabular}
\label{tab:baselineEAV}
\end{table}


Generally, the results indicate that the economic performance of CES is better than HES systems. While the scaling effect has a positive impact on the cost side, the stacking of multiple applications contributes positively to the benefit side of CES systems resulting in a total improvement of the profitability of CES over HES systems. Measured in terms of EAV, an improvement of 42\% on average can be achieved considering combinations in which both a PV and a BES system is present (avg. HES: -185 \EUR{} versus avg. CES: -108 \EUR{}). This is also supported by the additional stack of services. As shown in Table~\ref{tab:baselineUR}, even in terms of multiple utilization, the CES systems reach the peak at a UR of 63\%. On average, the HES is utilized by 44\% and the CES by 61\%. 

\begin{table}[H]
\centering
\caption{Comparison of aggregated form of HES and CES systems (operator and prosumer results) based on UR (the CES size[kWh] is given per household)}
\begin{tabular}{@{}llrrrrr@{}}
\toprule
& \multicolumn{6}{c}{\qquad HES [$kWh$]}      \\ \cmidrule{3-7}
\multirow{9}{*}{\rotatebox[origin=c]{90}{~PV [$m^2$]}} &    & \multicolumn{1}{c}{0} & \multicolumn{1}{c}{2.5} & \multicolumn{1}{c}{5} & \multicolumn{1}{c}{7.5} & \multicolumn{1}{c}{10} \\ \cmidrule{2-7}
& 0  & -   &  \multicolumn{1}{c}{-}   & \multicolumn{1}{c}{-}  &   \multicolumn{1}{c}{-}  &   \multicolumn{1}{c}{-}  \\\cmidrule{2-7}
 & 7.5 & -    & 27\%    &  30\% &   30\%   &   30\%  \\\cmidrule{2-7}
& 15   &  -   & 36\%     &  46\%  &  49\%   & 50\%   \\\cmidrule{2-7}
& 22.5 & -    & 37\%     & 51\%  &  56\%   &   57\% \\\cmidrule{2-7}
& 30  & -   & 37\%    & 53\%  & 59\%    & 60\%  \\ \bottomrule
\end{tabular}
\qquad \quad
\begin{tabular}{@{}llrrrrr@{}}
\toprule
& \multicolumn{6}{c}{\qquad CES [$kWh$]}      \\ \cmidrule{3-7}
\multirow{9}{*}{\rotatebox[origin=c]{90}{~PV [$m^2$]}} &    & \multicolumn{1}{c}{0} & \multicolumn{1}{c}{2.5} & \multicolumn{1}{c}{5} & \multicolumn{1}{c}{7.5} & \multicolumn{1}{c}{10} \\ \cmidrule{2-7}
& 0  & -   &   53\%    & 53\%    &   53\%    &   53\%   \\\cmidrule{2-7}
& 7.5 & -   &   59\%   & 60\%   &   60\%   &   60\%  \\\cmidrule{2-7}
& 15   & -   &   60\%   & 62\%   &   63\%   &   63\%  \\\cmidrule{2-7}
& 22.5   & -   &  60\%   & 62\%   &   63\%   &   63\%  \\\cmidrule{2-7}
& 30  & -   &  59\%   & 62\%   &  63\%   &   63\%   \\ \bottomrule
\end{tabular}
\label{tab:baselineUR}
\end{table}

The economic results of scenario 1 are summarized in Table \ref{tab:baselinesummary} with the help of the mean value of the 16 different PV and CES system combinations, excluding the sensitivities without CES or PV system capacities. While the results for the HES and CES cases have been presented in this section, the results of CES\textsubscript{OPT} and  CES\textsubscript{DIR} are explained and examined in the following Section \ref{sec:performanceanalysis}. The coefficient of variation c\textsubscript{v}\footnote{c\textsubscript{v} is defined as the standard deviation $\sigma$ to the mean $\mu$: \quad $c\textsubscript{v}= \frac{\sigma}{\mu}$ } constitutes a measure of relative variations \cite{KLINGLER20181064} to outline the level of variation of the 16 sensitivities on a percentage basis. The higher the c\textsubscript{v}, the greater is the relative degree of dispersion around the mean, which indicates that the result is highly sensitive to the system combination. 

\subsubsection{Reduction analysis}
\label{sec:performanceanalysis}

\begin{figure*}[h!]
	\centering
    \subfloat[CES capacity 2,5~kWh per household, PV size 7,5~m$^2$]{\includegraphics[width=\linewidth, height=\textheight,keepaspectratio]			           {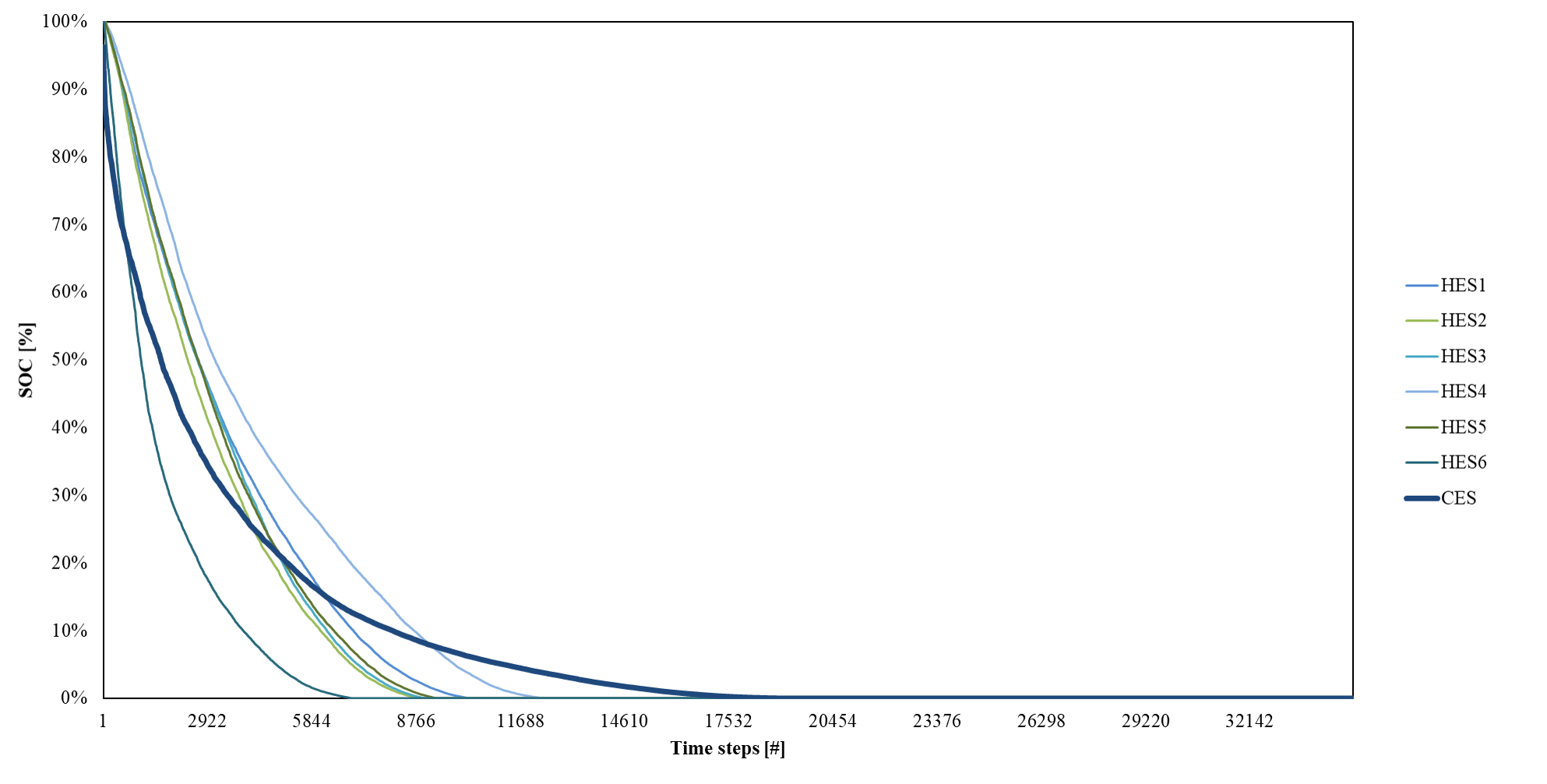}\label{fig:duration_curve_A}} \\
    \subfloat[CES capacity 5~kWh per household, PV size 7,5~m$^2$]{\includegraphics[width=\linewidth, height=\textheight,keepaspectratio]
   	{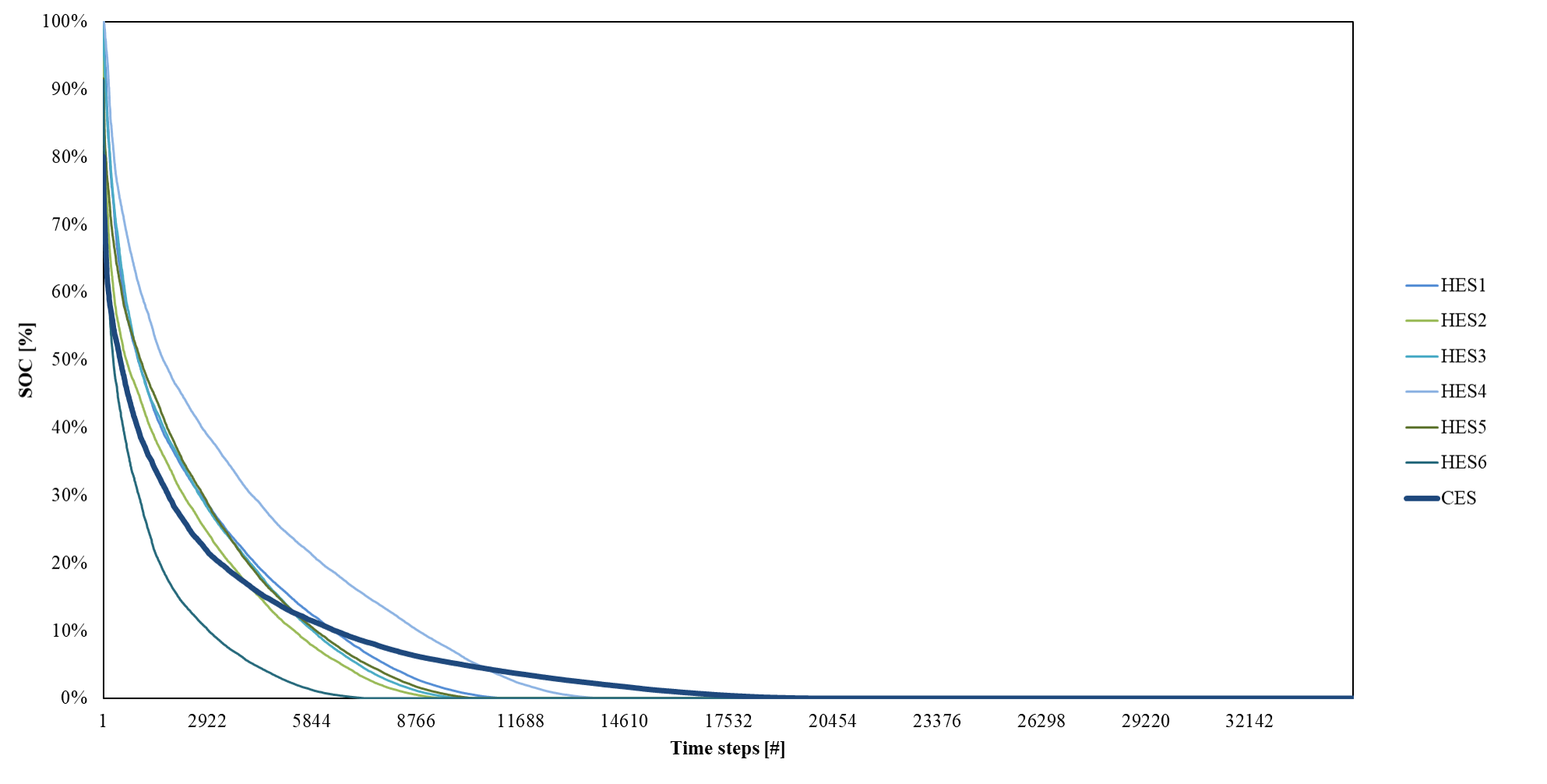}\label{fig:duration_curve_B}} \\
\centering
\caption{State of charge duration curves of the accumulated CES system compared to individual HES systems for selected sensitivities of the base case scenario 1}
\label{fig:duration_curve}
\captionsetup[subfigure]{labelformat=empty}    
\end{figure*}
As has been frequently noted, the optimal capacity with a CES system can be smaller than the aggregated capacities of all individual HES systems. To get first insights into the proper dimensioning of the CES system, it is helpful to identify the number of time steps in which the state of charge (SOC) reaches a certain percentage. Figure \ref{fig:duration_curve} presents two exemplary sorted annual SOC duration curves of two selected sensitivity cases of scenario 1. In this context, each Figure illustrates the variation of the SOC data of the CES system and HES systems in descending order of magnitude, rather than chronologically. While in Figure \ref{fig:duration_curve_A} each household is equipped with a 7.5 m\textsuperscript{2} PV system combined with a 2.5 kWh CES system share, in Figure \ref{fig:duration_curve_B} each household is equipped with a 7.5 m\textsuperscript{2} PV system combined with a 5 kWh CES system share.


In both figures, it is visible that the SOC duration curve of the CES system is steeper at the beginning and flatter at the end compared to the SOC duration curves of the HES systems. Concerning this matter, the SOC of the CES system utilizes the full storage capacity in only a few time steps (see Figure \ref{fig:duration_curve_A}) or not at all (see Figure \ref{fig:duration_curve_B}). Thus, a capacity reduction of the CES system might but does not necessarily have to result in an equivalent utilization restriction. In order to not place any household in a less favourable position regarding the MSC, the optimal reduction potential describes the capacity reduction of the CES at which it is still possible to utilize the CES in an optimal way as calculated by IRPopt. Thus, the downsized CES is also referred to as CES\textsubscript{OPT} as the adjusted CES is still sufficient to meet 100\% of the initial SC.

\begin{figure*}[h!]
	\centering
    \subfloat[CES capacity 2,5~kWh per household, PV size 7,5~m$^2$]{\includegraphics[width=\linewidth, height=\textheight,keepaspectratio]			           {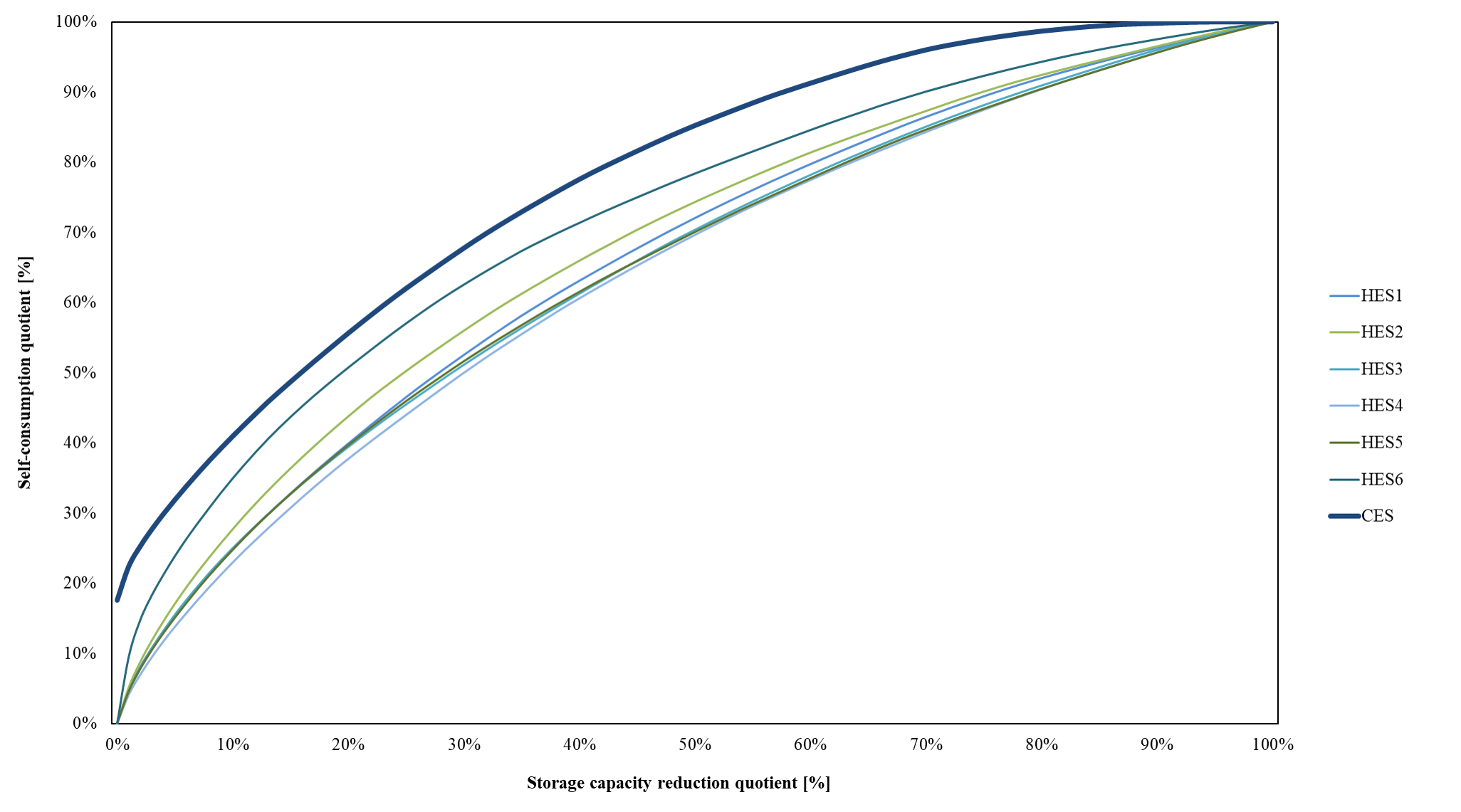}\label{fig:reduction_curve_A}} \\
    \subfloat[CES capacity 5~kWh per household, PV size 7,5~m$^2$]{\includegraphics[width=\linewidth, height=\textheight,keepaspectratio]
   	{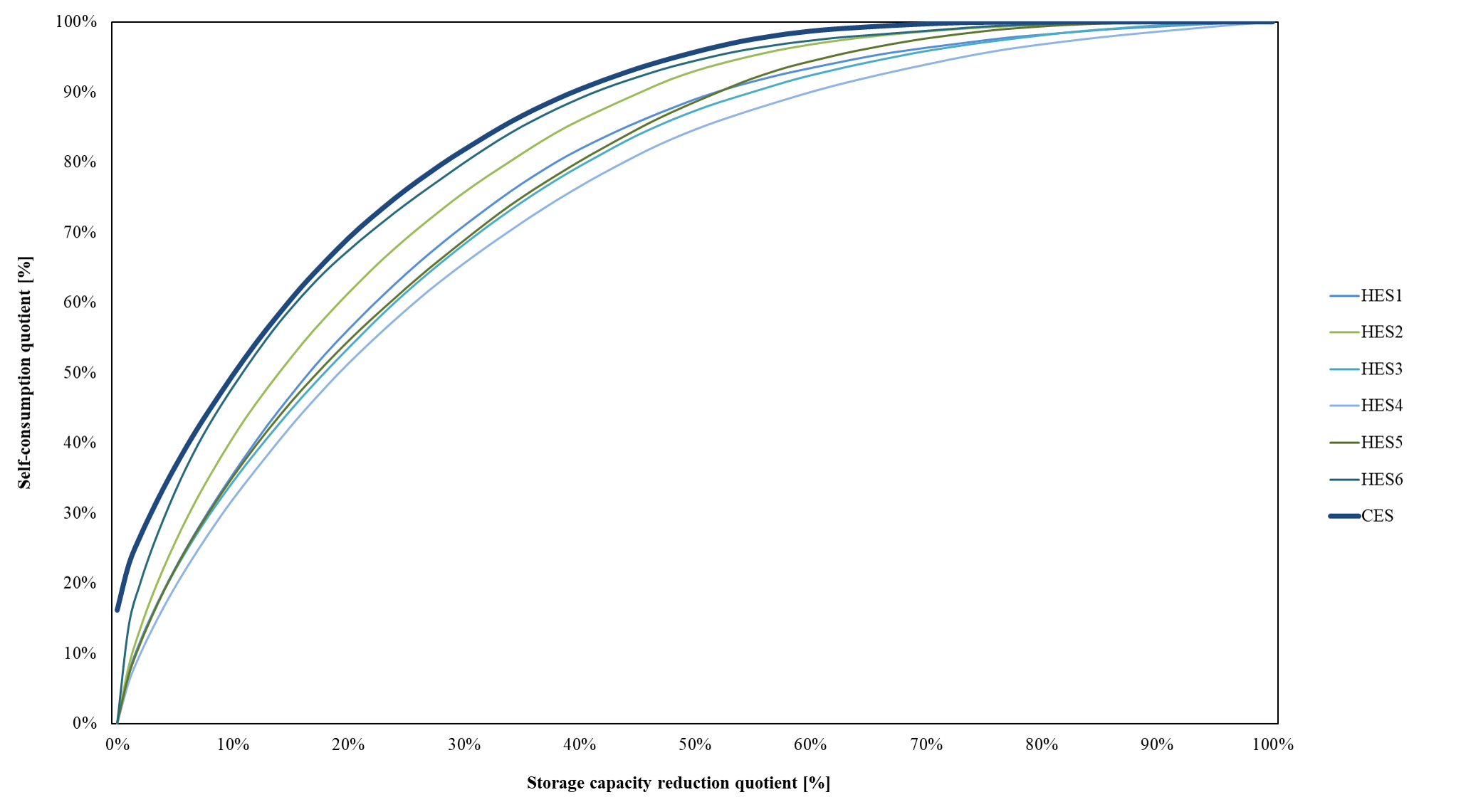}\label{fig:reduction_curve_B}} \\
\centering
\caption{Reduction potential of the accumulated CES system compared to individual HES systems depending on the self-consumable energy for selected sensitivities of the base case scenario 1}
\label{fig:reduction}
\end{figure*}

A deeper assessment of CES\textsubscript{OPT} requires the identification of the reduction potential of the storage capacity by taking into account the shortage of the $SCR_{tot}$ of the storage system. In this case, a reduction is only possible if the household is not cut with respect to the self-consumable energy. Thereby, only the rate representing the proportion of electricity which is stored in the CES ($E_{CES}$) to the total electricity generated ($E_{PV}$) is important. This has been defined as $SCR_{CES}$ in Section \ref{sec:keyindicators}. Thereby, only a high frequency of changes between very high and very low states of charge leads to a respective reduction of the $SCR_{CES}$. Figure~\ref{fig:reduction} depicts a characteristic curve of the self-consumption quotient linked to the storage capacity reduction quotient. The self-consumption quotient represents the ratio of the self-consumable energy in terms of the reduced storage capacity and the self-consumable energy in terms of the initial storage capacity. The storage reduction quotient states the percentage share of the storage capacity of the initial storage capacity. In other words, the determined annual self-consumed electricity of a selected sensitivity case is given by $SCR_{CES}$ which is normalized to 100\%. Besides, the chosen storage capacity is also shown in a normalized form with 100\%. 

If the CES system remains the chosen capacity (100\% of the storage capacity reduction quotient) the self-consumable energy stays the same (100\% of the self-consumption quotient) for both the HES systems and CES systems. This is show in Figures \ref{fig:reduction_curve_A} and \ref{fig:reduction_curve_B}. The other way round, if the capacity is reduced to zero and no capacity for the storage is left (0\% for the capacity reduction quotient), the self-consumable energy of the HES systems decreases to zero. For the cases with CES, however, the self-consumable energy only drops to CES\textsubscript{DIR} 18\% (see Figure \ref{fig:reduction_curve_A}) or CES\textsubscript{DIR}
 16\% (see Figure \ref{fig:reduction_curve_B}) of the initial value of $SCR_{CES}$. This is justified by the harmonization effect due to slightly varying demand and generation profiles. The energy could be directly used by other households of the community without a need to be stored at all (CES\textsubscript{DIR}). 

Moreover, as shown in both figures, the relative state of charge of the CES varies with a lower frequency and a smaller amplitude compared to the state of charge of the HES systems. Due to these two aspects the self-consumption as a percentage of $SCR_{CES}$ falls less strongly in terms of a capacity reduction of the CES. Thus, in contrast to HES systems, a moderate CES system reduction has a far smaller effect on the customer benefit regarding self-consumption. In concrete terms, the sensitivities of Figure \ref{fig:reduction_curve_A} allow a reduction of 4\% of the CES capacity without cutting $SCR_{CES}$. In contrast, the sensitivities of Figure \ref{fig:reduction_curve_B} enable a reduction of 20\%. Across the 16 PV and CES combinations, excluding again the sensitivities without a CES or PV system, the average reduction potential (CES\textsubscript{OPT}) is 8\% with a coefficient of variation of 1.99. Thus, on average the initial storage capacity can be reduced to 
92\%. This average reduction potential is utilized to determine lower specific investments of CES systems. The results are summarized in Table \ref{tab:baselinesummary}. There is a significant degree of dispersion around the mean and the results are quite sensitive. The c\textsubscript{v} of 0.18 ($\sigma$=16\%; $\mu$=92\%) is obtained for the resizing potential of CES\textsubscript{OPT} across the 16 combinations. 

\begin{table}[h!]
\centering

\caption{Mean AFB, EAC, EAV values across the sensitivities of scenario 1 and the corresponding coefficient of variation c\textsubscript{v} comparing HES, CES, CES\textsubscript{OPT} and CES\textsubscript{DIR}}
\begin{tabular}{@{} lrccrccrccrcc@{}}
\toprule
 & \multicolumn{2}{l}{HES} &  & \multicolumn{2}{l}{CES}&  & \multicolumn{2}{l}{CES\textsubscript{OPT}} && \multicolumn{2}{l}{CES\textsubscript{DIR}}&  \\ 
 \cmidrule{2-3} \cmidrule{5-6} \cmidrule{8-9} \cmidrule{11-12} 
 & mean & c\textsubscript{v} &  & mean & c\textsubscript{v} &  & mean & c\textsubscript{v} &  & mean & c\textsubscript{v} \\ \midrule
Storage size (of initial value)&100\%&-&&100\%&-&&92\%&0.18&&11\%&0.30  \\\midrule
AFB&122 \EUR{}&0.42&&157 \EUR{}&0.36&&151 \EUR{}&0.40 &&-&- \\ 
EAC&307 \EUR{}& 0.45&&265 \EUR{} &0.39&&241 \EUR{}&0.40&&-&-  \\
EAV& -185 \EUR{} & 0.68 &&-108 \EUR{} &0.63&&-90 \EUR{}&0.54&&-&-  \\
 \bottomrule
\end{tabular}
\label{tab:baselinesummary}
\end{table}

\subsubsection{Scenario overview}
\label{sec:6-3}
Table \ref{tab:eccomparisonscenarios} shows the economic results of CES systems. It can be observed that all other scenarios have a negative impact. The AFB is reduced in each scenario compared to the baseline scenario, and the CES system in scenario 4 achieves the lowest AFB. Compared to the base case, the drawbacks of the simultaneous application of flexibility options can reach up to 38\%. This also leads to the worst economic result with an EAV of -169 \EUR{}. Thereby, HP (scenario 2) affects the economics of CES systems in a stronger negative way than DR (scenario 3) while reduction of AFB caused by a lower P/E ratio is not that high.

\begin{table}[H]
\centering
\caption{Mean AFB, EAC and EAV values of the regular CES systems across the sensitivities of scenario 1-5 as well as the sample's coefficient of variation c\textsubscript{v} in different scenarios}
\resizebox{\columnwidth}{!}{%
\begin{tabular}{@{} lrccrccrccrccrcc@{}}
\toprule
 & \multicolumn{2}{c}{Scenario 1} &  & \multicolumn{2}{c}{Scenario 2}&  & \multicolumn{2}{c}{Scenario 3} && \multicolumn{2}{c}{Scenario 4}&&
 \multicolumn{2}{c}{Scenario 5}
 \\ 
 \cmidrule{2-3} \cmidrule{5-6} \cmidrule{8-9} \cmidrule{11-12} \cmidrule{14-15} 
 & mean & c\textsubscript{v} &  & mean & c\textsubscript{v} &  & mean & c\textsubscript{v} &  & mean & c\textsubscript{v} &  & mean & c\textsubscript{v}\\ \midrule
AFB&157 \EUR{} & 0.36&&102 \EUR{} & 0.42 && 141 \EUR{}&0.36&&96 \EUR{}&0.41&&152 \EUR{}&0.35 \\
EAC& 265 \EUR{}& 0.39&& 265 \EUR{}& 0.39 && 265 \EUR{}& 0.39&&265 \EUR{}& 0.39&&265 \EUR{}& 0.39 \\ \midrule
EAV&-108 \EUR{}& 0.63&&  -163 \EUR{}&0.45&& -124 \EUR{}& 0.57&&-169\EUR{}&0.44&&-113 \EUR{} & 0.62  \\
 \bottomrule
\end{tabular}
}
\label{tab:eccomparisonscenarios}
\end{table}

Table \ref{tab:urssr} summarizes the performance indicators. The SSR results achieved in the different scenarios are strikingly similar (52-53\%) except for scenario 2 and 4 with HPs (39\%). In scenarios 1, 3, and 5, the end users of the community are able to cover more than half of their electricity demand through the PV system and also with help of the CES system. In scenario 2, the level of self-sufficiency is considerably lower. This is due to the fact that a significant share of the PV-generated electricity is used for heat generation via the HP. The UR of the CES systems scarcely differ in the different scenarios. It can be also noted that the UR is very insensitive regarding different PV and CES combinations as the c\textsubscript{v} is very low.

\begin{table}[H]
\centering
\caption{Mean SSR and UR values across the sensitivities of scenario 1-5 and the sample's coefficient of variation c\textsubscript{v} in different scenarios}
\resizebox{\columnwidth}{!}{%
\begin{tabular}{@{} lrccrccrccrccrcc@{}}
\toprule
 & \multicolumn{2}{c}{Scenario 1} &  & \multicolumn{2}{c}{Scenario 2}&  & \multicolumn{2}{c}{Scenario 3} && \multicolumn{2}{c}{Scenario 4}&&
 \multicolumn{2}{c}{Scenario 5}
 \\ 
 \cmidrule{2-3} \cmidrule{5-6} \cmidrule{8-9} \cmidrule{11-12} \cmidrule{14-15} 
 & mean & c\textsubscript{v} &  & mean & c\textsubscript{v} &  & mean & c\textsubscript{v} &  & mean & c\textsubscript{v} &  & mean & c\textsubscript{v}\\ \midrule
SSR&52\% & 0.27&& 39\% & 0.33 && 53\%&0.28&&39\%&0.35&&52\%&0.25 \\
UR& 61\% & 0.02&& 58\%& 0.04&& 61\%& 0.02&&57\%& 0.04&&68\%& 0.02\\ 
 \bottomrule
\end{tabular}
}
\label{tab:urssr}
\end{table}

\begin{table}[H]
\centering
\caption{Mean absolute SC values, SCR and reduction potential across the sensitivities of scenario 1-5 as well as the sample's coefficient of variation c\textsubscript{v}}
\resizebox{\columnwidth}{!}{%
\begin{tabular}{@{} lrccrccrccrccrcc@{}}
\toprule
 & \multicolumn{2}{c}{Scenario 1} &  & \multicolumn{2}{c}{Scenario 2}&  & \multicolumn{2}{c}{Scenario 3} && \multicolumn{2}{c}{Scenario 4}&&
 \multicolumn{2}{c}{Scenario 5}
 \\ 
 \cmidrule{2-3} \cmidrule{5-6} \cmidrule{8-9} \cmidrule{11-12} \cmidrule{14-15} 
 & mean & c\textsubscript{v} &  & mean & c\textsubscript{v} &  & mean & c\textsubscript{v} &  & mean & c\textsubscript{v}  &  & mean & c\textsubscript{v}\\ \midrule
SC\textsubscript{tot} (kWh)&1,944&0.28&&2,547&0.36  && 2,004&0.28&&2,571&0.36&& 1,952 &0.26 \\
SC\textsubscript{CES} (kWh)&807&0.42&&338&0.68 &&661&0.44&&266&0.70&& 808 & 0.39  \\ \midrule
SCR\textsubscript{tot}&72\%&0.23&& 86\%&0.14  &&74\%&0.21&&90\%&0.10&&72\%&0.23 \\
SCR\textsubscript{CES}& 28\%&0.20&& 13\%&0.53  &&22\%&0.19&&7\%&0.48&& 28\% & 0.21 \\
SCR\textsubscript{EL}& 44\%&0.32&& 40\%&0.30 &&51\%&0.30&&41\%&0.23&& 44\%&0.30\\
SCR\textsubscript{HP}& -&-&& 33\%&0.37 &&-&-&&41\%&0.08&&-&-\\ \midrule
Reduction  \\
potential CES\textsubscript{OPT}&8\%&1.99&&19\%&1.30 &&11\%&1.82&&23\%&1.16&& 9\% &1.96\\ \midrule
SC with a zero \\
capacity CES\textsubscript{DIR} &11\%&0.30&&6\%&0.42 &&11\%&0.33&&5\%&0.55&& 11\% &0.28 \\
 \bottomrule
\end{tabular}
}
\label{tab:techcomparisonscenarios}
\end{table}

Table \ref{tab:techcomparisonscenarios} lists further meaningful performance parameters, which allows a comparison based on SC, SCR as well as the reduction potential of CES systems over all scenarios. Furthermore, Table \ref{tab:techcomparisonscenarios} shows the share of the initially determined SC, which can be achieved with a CES capacity of zero (CES\textsubscript{DIR}). The results show that the dimensioning of CES\textsubscript{OPT} systems is significantly influenced by the scenario configuration. In particular, interdependencies between systems and local flexibility options with consequences for the capacity can be observed. While the CES\textsubscript{OPT} can be reduced by 19\% in the presence of on-site HPs, it can be even reduced by 23\% if both DR and HP flexibilities are available for the households. The scenario assumptions in scenario 1 and scenario 5 allow for CES\textsubscript{OPT} capacity reduction of 8\%. The bottom line of Table \ref{tab:techcomparisonscenarios} displays the percentage of self-consumed electricity of the initially determined SC\textsubscript{CES}, which can be still achieved with a CES capacity of zero in each scenario (CES\textsubscript{DIR}).

\subsection{Investment assessment}
\begin{figure}[h]
\centering
\includegraphics[width=\linewidth, height=\textheight,keepaspectratio]{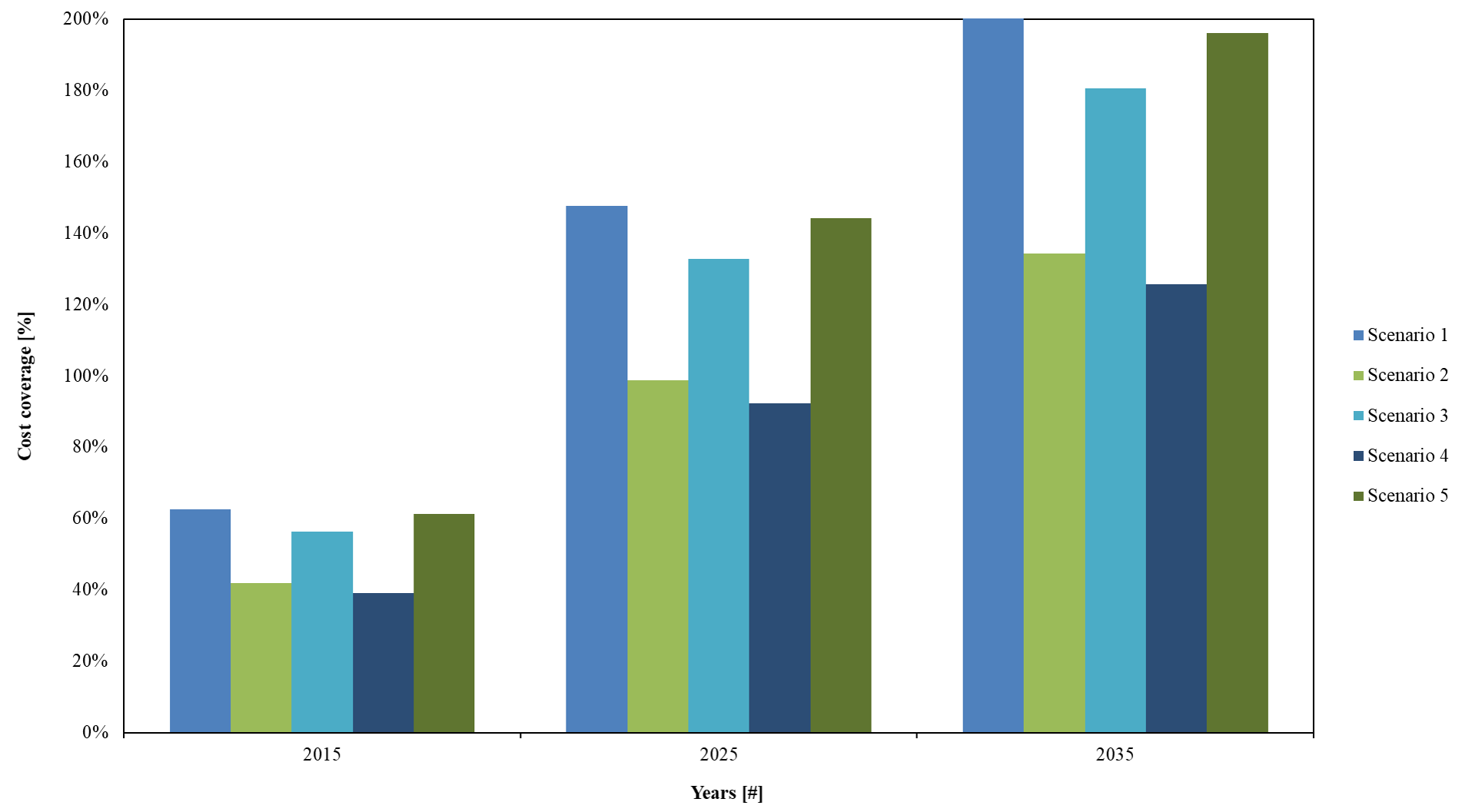}
\caption{Cost coverage of CES\textsubscript{OPT} of different scenarios for the years 2015, 2025, and 2035}
\label{fig_coverage}
\end{figure}

By taking into account the investment benefits of the CES capacity reduction as introduced in section \ref{sec:performanceanalysis} and summarized in \ref{sec:6-3}, as well as the drawbacks of the flexibility losses as introduced in section \ref{sec:6-1} and summarized in \ref{sec:6-3}, the optimization results allow conclusions to be drawn about the viability of the optimal CES model for the current and future years. Even though a cost reduction of about 14\% for the CES compared to the HES scenarios could be reached on average in the base case scenario 1, the EAV as outlined in Table \ref{tab:baselineEAV} is still negative in the year 2015. At the same time, by considering the average capacity reduction potential of 8\% in scenario 1 as shown in Table \ref{tab:techcomparisonscenarios}, further savings regarding EAC can be achieved. Even though this improves the profitability, the EAV balance of the CES\textsubscript{OPT} remains negative, as visible in Figure \ref{fig_coverage} in the year 2015. Additionally, for the years 2025 and 2035 the projected price developments for stationary Li-ion batteries, as investigated in Figure \ref{fig:future cost}, are taken into account for the investment assessment since they also positively affect the EAC.

In this context, the coverage ratio in Figure \ref{fig_coverage} depicts the average AFB and EAC across the 16 PV and CES sensitivity combinations of each scenario for the determined CES\textsubscript{OPT}. Within this Figure, 100\% indicates the break-even point at which the EAC and AFB is equal and thus the EAV is equal to zero. In terms of the applied scenarios 1-5 for the year 2015, the base case scenario 1 always represents the best case scenario for each year. This indicates that the further capacity reduction possibilities in the scenarios 2-5 and the corresponding investment savings (higher capacity reduction leads to smaller EAC) are outweighed by the drawbacks of cross-sectoral flexibility losses (multiple flexibility options lead to a smaller AFB).  This is also visible in Table \ref{tab:eccomparisonscenarios} and Table \ref{tab:techcomparisonscenarios} in more detail. In this context, especially the impact of a simultaneous application of a HP in scenario 2 and 4 has a higher impact on the AFB than on the EAC. A smaller power-to-energy ratio has a small negative influence. 

Moreover, while in the year 2015 none of the scenarios is a viable business case, in the year 2025 more than half of the scenarios are viable. Scenarios with the simultaneous application of a HP (scenarios 2 and 4) are still below the break even in 2025. In contrast, in the year 2035 all scenarios are viable. This can be mainly attributed to the investment reduction during the years 2015 to 2035. Thus, even in cases with a simultaneous application of one or more flexibility options (scenarios 2-4), the optimal sized CES reaches at least the profitability threshold through the years.

\subsection{Research implications}
\label{S:7}

The optimization results of the base case scenario are generally in line with the literature. CES systems with multiple applications are more cost-effective than HES systems with a single application as also shown by \cite{PARRA2017aop,PARRA2015,PARRA2016}. On average, systems with multiple applications are economically 42\% better off. There are essentially two reasons: first, the scaling  effect of CES systems reduces the relative investment compared to HES systems and second, adding operator applications increases total utilization and hence the revenues attained by the CES. Additionally, the aggregation of generation and demand profiles, which come into effect with shared CES systems, allow a relative reduction of the storage capacity. On average, reductions of 8-9\% are possible without any deterioration of SCR in the base case. The resized CES\textsubscript{OPT} improved the economics of the system by 16\%. Thereby, capacity reductions of 35\% as shown by \cite{BARBOUR2018}, could not be reached. The results of this paper are more in line with \cite{VANDERSTELT2018266}. 

Regarding the flexibility competition effects, it needs to be stated that the simultaneous application of flexibility options and especially cross-sectoral flexibilities have a negative impact on the profitability of CES systems. The benefits regarding investments savings are smaller than the drawbacks regarding flexibility losses. It was observed that HPs reduced the annually obtained benefits of CES by 35\% due to less utilization by end users. Even though, CES\textsubscript{OPT} systems have a higher capacity reduction potential of 19\%, the financial losses regarding the flexibility potential outweigh the cost savings regarding the capacity reduction. DR has a similar effect but not to the same extent. The simultaneous application of both flexibilities leads to even worse results. Finally, it was shown that a lower P/E ratio only slightly worsens the performance. In this context, only an integrated analysis of demand response, supply management, sector coupling, and energy storage solutions allows a comprehensive decision of complex models as also encouraged by \cite{hansen2019status}.

In terms of the storage market viability, no viable CES model could be determined for the year 2015. This has been calculated by taking the optimal capacity, flexibility potential, economies of scale and the cost trends of the storage capacities into account. The picture completely changes for the projected years from 2025 onwards. Despite the drawbacks of the simultaneous application of flexibility options most the CES scenarios are viable. This goes in line with the evaluated literature \cite{TERLOUW2019356}, which gives no precise information regarding the year in which the investment in a CES is going to pay off. Either, they are already economical \cite{TERLOUW2019356,ELBATAWI8038034}, or they will pay off in a zero-carbon year which is not further specified \cite{PARRA2015}. In 2035, the present paper suggests that the contribution margins of CES systems are positive in each investment scenario and the overall profitability of CES is higher again. This also shows that the economies of scale but also the cost trend development have positive effects on the investment decision as also shown by \cite{graditi2016technical}. The simultaneous application of flexibilities is reducing the viability of CES systems. Nevertheless, the assumed cost reduction has a stronger effect and CES seems to be a viable community solution for the next years.

All in all, communities seem to become an attractive scale for a successful energy transition. On the one hand, this paper shows an economic deployment of CES systems. On the other hand, it has also been shown that community models as direct marketing also can support the integration of renewable energy. The cost trends which refer to the investment of the CES are subject to a high degree of uncertainty. There have been differences with respect to future cost. Furthermore, the same storage lifetime has been assumed for HES and CES. Different studies state a lifetime of only 10 \cite{Reedman2015}, 20 \cite{MUNOZ2016,IRENA2017}, or even 30 years \cite{BVES2016}. However, a CES is more utilized compared to a HES. In this way, the results of the CES are possibly overestimated. Since the main contribution of this paper aims at comparing the CES scenarios among themselves, the implications for the results are almost negligible. Even though the lifetime of a CES would be a little bit higher in scenarios with competing flexibility options, the general statement about the degradation seems to be supported by the relatively similar UR of the CES in the different scenarios as shown in Table \ref{tab:urssr}. 

Moreover, the inclusion of more and diverse end-consumers might even lead to better results since the demand and generation could be balanced out in a better way. In the course of capacity scaling the specific investments would achieve an even better stance due to the economies of scale as also outlined by \cite{PARRA2015}. Additionally, variable tariffs would have a positive effect on the economic viability of CES systems. Thus, the analysis framework of this paper might not only be applicable to communities in Germany. However, favorable regulatory conditions as well as an enabling technical, policy and market environment are a prerequisite for the exploitation of the economic potential of CES systems \cite{Scheller.2018legal}. Finally, in contrast to HES a CES seem to provide various options for a community and also allows a quite flexible expansion or dismantling.

\section{Concluding remarks}
\label{S:8}

Community electricity storage systems are often seen as a key solution for the challenges raised by the ongoing energy system transition and are expected to offer promising potential across different actors along the energy system value chain. On the one hand, the competition effects for community electricity storage are analyzed, taking into account simultaneous flexibility options. On the other hand, the market viability is analyzed, taking into account elaborated storage cost trends. The results of the case studies provide support that the scaling effect, multi-use and the aggregated demand significantly enhance the value proposition of community storage systems compared to individual household storage systems. In the base case, storage capacity reductions of 9\% per household are possible. The simultaneous application of flexibility options enable a further capacity reduction of the community storage size by up to 23\%. However, the competition of flexibility options lead also to storage flexibility potential reductions from 11\% to 38\%. In total, the results show that the losses of the flexibility benefits outweigh the savings of the capacity investment reductions. Furthermore, the flexibility value achieved is not sufficient to reach breakeven at current average Li-ion battery costs. This seems to change between the years of 2025 and 2035 as the investments are expected to decrease. Whether and to what extent the flexibility revenues might change regarding the spot market and balancing market is uncertain. 

In practice, a comprehensive integrated analysis is required, taking into account not only the storage application itself but also the behavior of the participating customers. This would also raise the question regarding the prioritization, of ownership models and the application prioritization since the willingness to invest in such a technology is linked to the individual revenues of each participant. The presented case studies demonstrate that the model IRPopt constitutes a valuable tool in order to assess the complexities of community models. Future research should further investigate the simultaneous performance of several applications such as spot market trading, peak shaving, demand response and balancing energy services to enhance the revenue potential of community storage systems as already partially done for industrial customers in \cite{braeuer2019battery}. Especially the question of which applications are competing and which may complement each other should be answered in detail. In this context, various policy frameworks to support demand-side flexibility options should also be assessed.

\section*{Declarations of interest} 
There are no competing interests.

\section*{Acknowledgement}
Fabian Scheller receives funding from the European Union's Horizon 2020 research and innovation programme under the Marie Sklodowska-Curie grant agreement no. 713683 (COFUNDfellowsDTU). Russell McKenna kindly acknowledges the financial support support of the projects Smart City Accelerator and FlexSUS: Flexibility for Smart Urban Energy Systems (Project No. 91352), which has received funding in the framework of the joint programming initiative ERA-Net Smart Energy Systems’ focus initiative Integrated, Regional Energy Systems. The usual disclaimer applies.


\bibliographystyle{elsarticle-num}
\bibliography{ModelDescription}



\end{document}